\documentclass[preprint]{JHEP3} % 10pt is ignored!

\JHEPspecialurl{http://jhep.sissa.it/JOURNAL/JHEP3.tar.gz}

\usepackage{epsfig,multicol,bbm}
\usepackage{citesort}
\newcommand\fverb{\setbox\fverbbox=\hbox\bgroup\verb}
\newcommand\fverbdo{\egroup\medskip\noindent%
\fbox{\unhbox\fverbbox}\ }
\newcommand\fverbit{\egroup\item[\fbox{\unhbox\fverbbox}]}
\newbox\fverbbox

\newcommand{\lsim}{\raisebox{-0.13cm}{~\shortstack{$<$ \\[-0.07cm] $\sim$}}~} 
\newcommand{\gsim}{\raisebox{-0.13cm}{~\shortstack{$>$ \\[-0.07cm] $\sim$}}~}

\newcommand{\beq}{\begin{eqnarray}} 
\newcommand{\eeq}{\end{eqnarray}} 
 
\preprint{LPT Orsay 10-15,\\ CERN-PH-TH/2010--051}

\title{Predictions for Higgs production at the Tevatron and the associated
uncertainties}

\author{Julien Baglio\\
Laboratoire de Physique Th\'eorique, Universit\'e Paris-Sud XI et CNRS,
F-91405 Orsay Cedex, France.\\
E-mail: \email{Julien.Baglio@th.u-psud.fr}}

\author{Abdelhak Djouadi\thanks{Permanent address: Laboratoire de Physique 
Th\'eorique, Unit\'e mixte CNRS et Universit\'e Paris-Sud XI, F-91405 Orsay Cedex, 
France.}\\
Theory Unit, CERN, 1211  Gen\`eve 23, Switzerland.\\
E-mail: \email{Abdelhak.Djouadi@cern.ch}}

\abstract{We update the theoretical predictions for the production cross
sections of the Standard Model Higgs boson at the Fermilab Tevatron collider,
focusing on the two main search channels, the gluon--gluon fusion mechanism
$gg\to H$ and the Higgs--strahlung processes $q \bar q \to VH$ with $V=W/Z$,
including all relevant higher order QCD and electroweak corrections in
perturbation theory.  We then estimate the various uncertainties affecting
these predictions: the scale uncertainties which are viewed as a measure of the
unknown higher order effects, the uncertainties from the parton distribution
functions and the related errors on the strong coupling constant, as well as
the uncertainties due to the use of an effective theory approach in the
determination of the radiative corrections in the $gg\to H$ process at
next-to-next-to-leading order. We find that while the cross sections are well
under control in the Higgs--strahlung processes, the theoretical uncertainties
are rather large in the case of the gluon--gluon fusion channel, possibly
shifting the central values of the next-to-next-to-leading order cross sections
by more than $\approx 40\%$. These uncertainties are thus significantly larger
than the $\approx 10\%$ error assumed by the CDF and D0 experiments in their
recent analysis that has excluded the Higgs mass range $M_{H}\!=\!162$--166 GeV
at the 95\% confidence level. These exclusion limits should be, therefore,
reconsidered in the light of these large theoretical uncertainties.}


\begin{document} 

\renewcommand{\thefootnote}{\arabic{footnote}}
\setcounter{footnote}{0}
\setcounter{page}{2}

\section{Introduction}

We are approaching the exciting and long awaited times of discovering the 
``Holy Grail"~of nowadays particle physics: the Higgs boson
\cite{Higgs,Review}, the remnant of the mechanism breaking the electroweak
gauge symmetry and at the origin of the particle masses. Indeed, the Large
Hadron Collider (LHC) has started to have its first collisions \cite{LHCnews}, 
although at energies and with instantaneous luminosities yet far from those
which would be required for discovery. Most importantly in this context, the
CDF and D0 experiments at the Fermilab Tevatron collider have collected enough
data to be sensitive to the Higgs particle of the Standard Model. Very
recently, the two collaborations performed a combined analysis on the search
for this particle and excluded at the 95\% confidence level the possibility of
a Higgs boson in the mass range between 162 and 166 GeV \cite{Tevatron0}; this
exclusion range is expected to increase to $159$ GeV  $\le M_H \le 168$ GeV
\cite{Tevatron}. We are thus entering a new era in the quest of the Higgs
particle as this is the first time that the mass range excluded by the LEP
collaborations in the late 1990s, $M_H \ge 114.4$ GeV \cite{LEP-Higgs}, is
extended.

However, in contrast to the Higgs LEP limit which is rather robust, as the
production cross section is mainly sensitive to small electroweak effects that
are well under control, the Tevatron exclusion limit critically depends on the
theoretical prediction for the Higgs production cross sections which, at hadron
colliders, are known to be plagued with various uncertainties. Among these are
the contributions of yet uncalculated higher order corrections  which can be
important as the strong coupling constant $\alpha_s$ is rather large, the errors
due to the folding of the partonic cross sections with the parton distribution
functions (PDFs) to obtain the production rates at the hadronic level, and the
errors on some important input parameters such as $\alpha_s$. It is then
mandatory to estimate these uncertainties in order to have a reliable
theoretical prediction  for the production rates, that would allow for a
consistent confrontation between theoretical results and experimental
measurements or exclusion bounds\footnote{An example of such a situation is  the
$p\bar p \to b\bar b$  production cross section that has been measured at the
Tevatron (and elsewhere) and which was a factor of two to three larger than the
theoretical prediction, before higher order effects and various uncertainties
were included. For a review, see Ref.~\cite{Foot0} for instance.}. The present
paper critically addresses this issue. 

At the Tevatron, only two production channels are important for the  Standard
Model Higgs boson\footnote{The CDF/D0 exclusion limits \cite{Tevatron} have
been obtained by considering a large variety of Higgs production and decay
channels (36 and 54 exclusive final states for, respectively, the CDF and D0
collaborations) and combining them using artificial neural network techniques.
However, as will be seen later, only a few channels play a significant role  in
practice.}. In the moderate to high mass range, 140 GeV$\lsim M_H \lsim 200$
GeV, the Higgs boson decays dominantly into $W$ boson pairs (with one $W$ state
being possibly off mass--shell) \cite{HDECAY} and the main production channel
is the gluon--gluon fusion mechanism  $gg \to H$ \cite{ggH-LO} which proceeds
through heavy (mainly top and, to a lesser extent, bottom) quark triangular
loops. The Higgs particle is then detected through the  leptonic decays of the
$W$ bosons, $H\to WW^{(*)}\to \ell^+ \nu \ell^- \bar \nu$ with $\ell=e,\mu$,
which exhibits different properties than the $p\bar p\to W^+ W^- \to \ell \ell$
plus missing energy continuum background \cite{HWW-DD}. 

It is well known that the $gg\to H$ production process  is subject to extremely
large QCD radiative corrections
\cite{ggH-NLO,SDGZ,Michael-rev,ggH-NNLO1,ggH-NNLO2,ggH-NNLO3,ggH-resum,ggH-FG,ggH-ADGSW}.
In contrast, the electroweak radiative corrections are much smaller, being at
the level of a few percent \cite{ggH-EW,ggH-actis,ggH-radja}, i.e.~as in the
case of Higgs production at the LEP collider.
For the corrections due to the strong interactions, the $K$--factor defined as
the ratio of the higher order (HO) to the lowest order (LO) cross 
sections, consistently evaluated with the $\alpha_s$ value and the PDF sets
at the chosen order, 
\beq
K_{\rm HO}= \sigma^{\rm HO} |_{ (\alpha_s^{\rm HO} \, , \, {\rm PDF^{HO} )} } 
\; / \; 
           \sigma^{\rm LO} |_{ (\alpha_s^{\rm LO} \, , \,  {\rm PDF^{LO}}) } \, , 
\label{kfactor}
\eeq
is about a factor of two at next-to-leading order  (NLO) \cite{ggH-NLO,SDGZ}
and about a factor of three at the next-to-next-to-leading  order (NNLO)
\cite{ggH-NNLO1,ggH-NNLO2,ggH-NNLO3}.  In fact, this exceptionally large $K$--factor is what allows
a sensitivity on the Higgs boson at the Tevatron with the presently collected
data. Nevertheless, the $K$--factor is so large that one may question the
reliability of the perturbative series, despite of the fact that there seems to be
kind of a convergence of the series as the NNLO correction is smaller
than the NLO correction\footnote{At LHC energies, the problem of the
convergence of the perturbative series is  less severe as the QCD
$K$--factor is only $\sim 1.7$ at NLO and $\sim 2$ at NNLO in the relevant
Higgs mass range.}. 

In the low mass range, $M_H \lsim 140$ GeV, the main Higgs decay channel is 
$H\to b\bar b$ \cite{HDECAY} and the $gg$ fusion mechanism cannot be used
anymore as the $gg \to H \to b\bar b$ signal is swamped by the huge QCD jet
background. The Higgs particle has then to be detected through its associated
production with a $W$ boson  $q \bar q \to WH$ \cite{HV-LO} which leads to
cleaner $\ell \nu b \bar b$ final states \cite{HV-SMW}. Additional topologies
that can also be considered in this context are  $q \bar q \to  WH$ with $H \to
WW^* \to \ell \ell \nu \nu$ or the twin production process    $q \bar q \to 
ZH$ with the subsequent decays $H \to b\bar b$ and $Z \to \nu \bar \nu$ or
$\ell ^+\ell^- $. Other production/decay channels are expected to lead to very
low rates and/or to be afflicted with too large QCD backgrounds. 

At the Tevatron, the   Higgs--strahlung processes $q \bar q \to VH$  with
$V=W,Z$ receive only moderate higher order corrections: the QCD corrections
increase the cross sections by about 40\% at NLO \cite{HV-NLO} and 10\% at NNLO
\cite{HV-NNLO}, while the impact of the one--loop electroweak corrections is 
small, leading to a $\approx 5\%$ decrease of the cross sections \cite{HV-EW}.
Thus, in contrast to the gluon--gluon fusion process, the production cross
sections in the Higgs--strahlung processes should be well under control. 

In this paper, we first update the cross sections for these two main Higgs
production channels at the Tevatron, including all known and relevant higher
order QCD and electroweak corrections and using the latest MSTW2008 set of 
parton distribution functions \cite{PDF-MSTW}. For the the $gg \to H$ process,
this update has been performed in various recent analyses
\cite{ggH-FG,ggH-radja} and, for instance, the normalized Higgs production
cross sections used by the CDF/D0 collaborations  in their combined analysis
\cite{Tevatron} are taken from these references. Such an update is lacking in
the case of the  Higgs--strahlung production channels $q\bar q \to VH$ and, for
instance, the normalised cross sections used by the Tevatron experiments
\cite{Tevatron} are those given in Ref.~\cite{HV-all} which make use of the old
MRST2002 set of PDFs \cite{PDF-MRST}, a  parametrisation that was approximate
as it did not include the full set of evolved PDFs at NNLO. For completeness,
we also update the cross sections for the two other single Higgs production
channels at hadron colliders: the weak boson fusion $p \bar p \to qqH$
\cite{VVH-LO,VVH-NLO} and the associated production with top quark pairs  $p
\bar p \to t \bar t H$ \cite{ttH-LO,ttH-NLO}. These channels play only a minor
role at the Tevatron but have also been included in the CDF/D0 analysis
\cite{Tevatron}. 

A second goal of the present paper is to investigate in a comprehensive way the
impact of all possible sources of uncertainties on the total cross sections for
the two main Higgs production channels. We first reanalyse the uncertainties
from the unknown higher order effects, which are usually estimated by exploring
the cross sections dependence on the renormalisation scale $\mu_R$ and the 
factorisation scale $\mu_F$. In most recent analyses, the two scales are varied
within a factor of two from a median scale which is considered as the most
natural one. We show that this choice slightly underestimates the higher order
effects and we use a criterion that allows a more reasonable estimate of the
latter: the range of variation of the two scales $\mu_R$ and $\mu_F$  should be
the one which allows the uncertainty band of the NLO cross section to match the
central value of the cross section at the highest calculated order. In the case
of $gg \to H$, for the uncertainty band of the NLO cross section to reach the
central result of the NNLO cross section, a variation of  $\mu_R$ and $\mu_F$
within a factor of $\sim 3$ from the central value $\mu_R= \mu_F =M_H$ is
required. When the scales are varied within the latter range, one obtains an
uncertainty on the NNLO cross section  of $\approx 20\%$, which is slightly
larger than what is usually assumed.

We then discuss the errors resulting from the folding of the partonic cross
sections with the parton densities, considering not only the recent MSTW set 
of PDFs as in Refs.~\cite{ggH-FG,ggH-ADGSW,ggH-radja}, but also two other PDF
sets that are available in the literature: CTEQ  \cite{PDF-CTEQ} and ABKM
\cite{PDF-Alekhin}. In the case of the cross section for the $gg \to H$ process
at the Tevatron, we find that while the PDF uncertainties evaluated within the
same scheme are moderate, as also shown in
Refs.~\cite{ggH-FG,ggH-ADGSW,ggH-radja}, the central values of the cross
sections obtained using the three schemes can be widely different. We show that
it is only when the experimental as well as the theoretical errors on the
strong coupling constant $\alpha_s$ are accounted for that one obtains results
that are consistent when using the MSTW/CTEQ and ABKM schemes. As a result, the
sum of the PDF+$\Delta^{\rm exp} \alpha_s$ and $\Delta^{\rm th} \alpha_s$ 
uncertainties, that we evaluate using a set--up recently proposed by the MSTW
collaboration to determine simultaneously the errors due to the PDFs and to
$\alpha_s$, is estimated to be at least a factor of two larger than what is
generally assumed.  

Finally, a third source of potential errors is considered in the $gg$
fusion mechanism: the one resulting from the use of an effective field theory
approach, in which the loop particle masses are assumed to be much larger than
the Higgs boson mass, to evaluate the NNLO contributions. While this error is
very small in the case of the top--quark contribution, it is at the percent
level in the case of the $b$--quark loop contribution at NNLO QCD where the
limit $M_H \ll m_b$ cannot be applied. This is also the case of the three--loop
mixed QCD--electroweak radiative corrections that have obtained in the
effective limit $M_H \ll M_W$, which lead to a few percent uncertainty. In
addition, an uncertainty of about 1\% originates from the freedom in the
choice  of the input $b$--quark mass in the $Hgg$ amplitude. The total
uncertainty in this context is thus not negligible and amounts to a few
percent.

We then address the important issue of how to combine the theoretical errors
originating from  these different sources. Since using the usually adopted
procedures of adding these errors either in quadrature, as is done  by the
experimental collaborations for instance, or linearly as is generally the case
for theoretical errors, lead to either an underestimate or to an overestimate
of the total error, we propose a procedure that is, in our opinion, more
adequate. One first determines the maximal and minimal values of the cross
sections obtained from the variation of the renormalisation and factorisation
scales, and then estimate directly on these extrema cross sections the 
combined uncertainties due to the PDFs and to the experimental and theoretical
errors on $\alpha_s$. The other smaller theoretical uncertainties, such as
those coming from the use of the effective approach in $gg\to H$, can be then
added linearly to this scale, PDF and $\alpha_s$ combined error.

The main result of our paper is that, when adding all these uncertainties using
our procedure, the total theoretical error on the production cross sections is
much larger than what is often quoted in the literature. In particular, in the
case of the most sensitive Higgs production channel at the Tevatron,  $gg \to H
\to \ell \ell \nu \nu$,  the overall uncertainty on the NNLO total cross
section is found to be of the order of  $\approx -40\%$ and $\approx + 50\%$.
This is significantly larger than the uncertainty of $\approx \pm 10\%$ assumed
in earlier studies and adopted in the CDF/D0 combined Higgs search analysis. As
a result, we believe that the exclusion range  given by the Tevatron
experiments for the Higgs mass in the Standard Model, 162 GeV $\le M_H \le 166$
GeV, should be reconsidered in the light of these results.  

The rest of the paper is organised as follows. In the next section we outline
our calculation of the Higgs production cross sections at the Tevatron in the
gluon--gluon fusion and Higgs--strahlung processes. In section 3, we focus on
the gluon--gluon fusion channel and evaluate the theoretical uncertainties  on
the cross section from scale variation, PDF and $\alpha_s$ uncertainties as
well as from the use of the effective theory approach for the NNLO
contributions. Section 4 addresses the same issues for the associated Higgs
production channels. The various theoretical errors are summarized and combined
in section 5 and their implications are discussed. A brief conclusion is given
in section 6.

\section{The production cross sections}

In this section, we summarize the procedure which allows to obtain our updated
central or ``best" values of the total cross sections for Higgs production at
the Tevatron in the Standard Model. We mainly discuss the two dominant
channels, namely  the gluon--gluon fusion and Higgs--strahlung,
but for completeness, we mention the two other production channels: vector
boson fusion and associated Higgs production with top quark pairs. 

The production rate for the $gg\to H+X$ process, where X denotes the additional
jets that appear at higher orders in QCD,  is evaluated in the following way.
The cross section up to NLO in QCD is calculated using the Fortran code {\tt
HIGLU} \cite{Higlu,Michael} which includes the complete set of radiative
corrections at this order, taking into account the full dependence on the top
and bottom quark masses \cite{SDGZ}. The contribution of the NNLO corrections
\cite{ggH-NNLO1,ggH-NNLO2,ggH-NNLO3}  is then implemented in this program using
the analytical expressions given in Ref.~\cite{ggH-NNLO2}. These corrections
have been derived  in an effective approach in which only the dominant top
quark contribution is included in the infinite top quark mass limit but the
cross section was rescaled by the exact $m_t$ dependent Born cross section, an
approximation which at NLO is accurate at the level of a few percent for Higgs
masses below the $t\bar t$  kinematical threshold, $M_H \lsim 300$ GeV
\cite{SDGZ,Michael-rev}. The dependence on the renormalisation scale $\mu_R$
and the factorisation scale  $\mu_F$ of the partonic NNLO cross sections  has
been reconstructed from the scale independent  expressions of
Ref.~\cite{ggH-NNLO2} using the fact that the full hadronic cross sections do
not depend on them and the $\alpha_s$ running between the $\mu_F$ and $\mu_R$
scales\footnote{The analytical expressions for the scale dependence have only
been given in  Ref.~\cite{ggH-NNLO3} in  the limit $\mu_F\!=\!\mu_R$ from which
one can straightforwardly obtain the case $\mu_F\!\neq \!\mu_R$ (see also
Ref.~\cite{ggH-NLO-resum}). We find agreement with this reference once the
virtual+soft $gg\!\to\!H$ partonic cross sections given in the Appendix are
multiplied by the factor $C_H$ given in eq.~(2.7). We thank V. Ravindran for
kindly clarifying this point to us.}. Nevertheless, for the central values of
the  cross sections which will be discussed in the present section, we adopt
the usual scale choice $\mu_R=\mu_F=M_H$. 

An important remark to be made at this stage is that we do not include the
soft--gluon resumation contributions which, for the total cross section, have
been calculated up to next-to-next-to-leading logarithm (NNLL) approximation
and increase the production rate by $\sim 10$--15\% at the Tevatron 
\cite{ggH-resum}. We also do not include the additional small contributions of
the estimated contribution at N$^3$LO \cite{ggH-N3LO} as well as those of  soft
terms beyond the NNLL approximation \cite{ggH-N3LO-others}. The reason is that
these corrections are known only for the inclusive total cross section and not
for the cross sections when experimental cuts are incorporated; this is also
the case for the differential cross sections \cite{ggH-cuts} and many
distributions that are used experimentally, which have been evaluated only at
NNLO at most. This choice of ignoring the contributions beyond NNLO\footnote{
One could also advocate the fact that it is theoretically not very consistent
to fold  a resumed cross section with PDF sets which do not involve any
resumation, as  is the case for the presently available PDF sets which at at
most at NNLO (although the effects of the resumation on the PDFs might be
rather small in practice); see for instance the discussion given in
Ref.~\cite{Magnea}.} has also been adopted  in Ref.~\cite{ggH-ADGSW} in which
the theoretical predictions have been confronted to the CDF/D0 results,  the
focus being the comparison between the distributions obtained from the matrix
elements  calculation with those given by the event generators and Monte-Carlo
programs used by the experiments. Nevertheless, the NNLL result for the cross
section can be very closely approached by evaluating the NNLO cross section at
the renormalisation and factorisation scales $\mu_R=\mu_F=\frac12 M_H$ 
\cite{ggH-resum} as will be commented upon later.

For the electroweak part, we include the complete one--loop  corrections to the
$gg\to H$ amplitude which have been calculated in Ref.~\cite{ggH-actis} taking
into account the full  dependence on the top/bottom quark and the $W/Z$ boson
masses. These corrections are implemented in the so--called partial
factorisation scheme in which the  electroweak correction $\delta_{\rm EW}$ is
simply added to the QCD corrected cross section at NNLO, $\sigma^{\rm tot}=
\sigma^{\rm NNLO} + \sigma^{\rm LO} (1+\delta_{\rm EW})$.  In the alternative
complete factorization scheme discussed  in Ref.~\cite{ggH-actis}, the
electroweak correction $1+\delta_{EW}$ is multiplied by the fully QCD corrected
cross section, $\sigma^{\rm tot}=\sigma^{\rm NNLO}(1+ \delta_{EW})$ and, thus,
formally involves  terms of  ${\cal O} (\alpha_s^3 \alpha)$ and ${\cal O}
(\alpha_s^4 \alpha)$ which have not been fully calculated. Since the QCD
$K$--factor is large, $K_{\rm NNLO} \approx 3$, the electroweak corrections
might be overestimated by the same factor. We have also included the mixed
QCD--electroweak corrections at NNLO due to light-quark loops \cite{ggH-radja}.
These are only part of the three--loop ${\cal O}(\alpha \alpha_s)$ corrections
and have been calculated in an effective approach that is valid only when $M_H
\lsim M_W$ and which cannot be easily extrapolated to $M_H$ values above this
threshold; this will be discussed in more details in the next section. In
Ref.~\cite{ggH-radja}, it has been pointed out that this procedure, i.e. adding
the NLO full result and the mixed QCD--electroweak correction  in the partial
factorization scheme, is equivalent to simply including only the NLO
electroweak correction in the complete factorisation scheme.

In the case of the $q\bar q \to WH$ and $q \bar q \to ZH$ associated Higgs 
production processes, we use the Fortran code {\tt V2HV} \cite{Michael} which 
evaluates the full cross sections at NLO in QCD. The NNLO QCD contributions to
the cross sections \cite{HV-NNLO},  if the $gg \to ZH$ contribution (that does
not appear in the case of $WH$ production and is at the permille level at the
Tevatron) is ignored, are the same as for the Drell--Yan process $p \bar p \to
V^*$ with $V=W,Z$ \cite{Drell-Yan} given in Ref.~\cite{DYNNLO,ggH-NNLO1}, once
the scales and the invariant mass of the final state are properly adapted.
These NNLO corrections, as well as the one--loop  electroweak corrections
evaluated in Ref.~\cite{HV-EW}, are incorporated  in the program {\tt V2HV}. 
The central scale adopted in this case is the invariant mass of the $HV$
system, $\mu_R=\mu_F = M_{HV}$.  

Folding the partonic cross sections with the most recent set of  MSTW parton
distribution functions \cite{PDF-MSTW} and setting the renormalisation and
factorisation scales at the most natural values discussed above, i.e.
$\mu_R\!=\!\mu_F\!=\!M_H$ for $gg \to H$ and  $\mu_R\!=\!\mu_F\!=\!M_{HV}$ for
$q\bar q \to VH$, we obtain for the Tevatron energy $\sqrt s=1.96$ TeV, the
central values  displayed in Fig.~1 for the Higgs production cross sections as
a function of the Higgs mass. Note that  we have corrected the numbers that we
obtained in an earlier version of the paper for the $p\bar p \to HW$ cross
section to include in the {\tt V2HV} program the CKM matrix elements  when
folding the partonic $q \bar q' \to HW$ cross sections with the parton 
luminosities\footnote{We thank R.  Harlander and Tom Zirke for pointing this 
problem to us.}; this results in a decrease of the $p\bar p \to  HW$ cross
section by $\approx 4\%$. In addition, it recently appeared that  including the
combined HERA data and the Tevatron $W\to \ell \nu$ charge asymmetry  data in
the MSTW2008 PDF set \cite{MSTW-new} might lead to an increase of the $p \bar 
p \to  (H+)Z/W$ cross sections by $\approx 3$\%; a small change in $\sigma
(gg \to H)$ is also expected.  

For the cross sections of the two sub-leading processes $qq \to V^* V^* qq \to
Hqq$ and $q \bar q/gg \to t\bar t H$ that we also include in Fig.~1 for
completeness, we have not entered into very sophisticated considerations. We
have simply followed the procedure outlined in Ref.~\cite{Review} and used the
public Fortran codes again given in Ref.~\cite{Michael}. The vector boson total
cross section is evaluated at NLO in QCD \cite{VVH-NLO} at a scale
$\mu_R=\mu_F=Q_V$ (where $Q_V$ is the momentum transfer at the gauge boson
leg), while the presumably small electroweak corrections, known for the LHC
\cite{VVH-EW}, are omitted. In the case of associated $t\bar tH$  production,
the LO cross section is evaluated at  scales $\mu_R=\mu_F=\frac12(M_H+2m_t)$
but is multiplied by a factor $K \sim 0.8$ over the entire Higgs mass range to
account for the bulk of the NLO QCD corrections \cite{ttH-NLO}. In the latter
case, we use the updated value $m_t=173.1$ GeV for the top quark mass
\cite{Top-mass}. The only other update compared to the cross section values
given in Ref.~\cite{Review} is thus the use of the recent MSTW set of PDFs.

\begin{figure}[!t]
\begin{center}
%\vspace*{-5mm}
\hspace*{-.5cm}
\epsfig{file=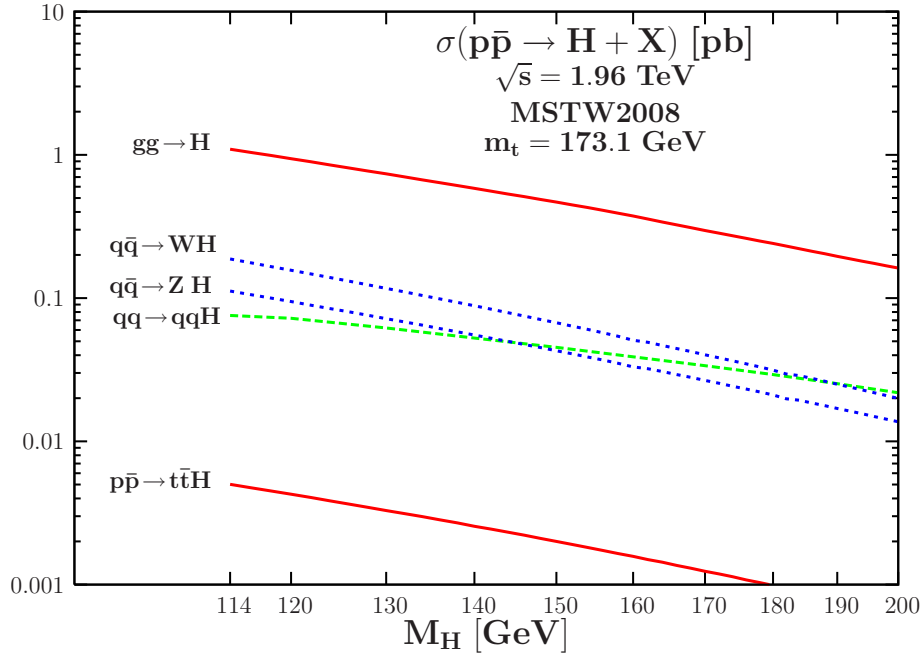,width=16.cm} 
\end{center}
\vspace*{-.5cm}
\caption[]{The total cross sections for Higgs production at the Tevatron 
as a function of the Higgs mass. The MSTW set of PDFs has been used and
the higher order corrections are included as discussed in the text.} 
\end{figure}

In the case of the $gg \to H$ process, our results for the total cross sections
are appro\-ximately 15\% lower than those given in
Refs.~\cite{Tevatron,ggH-FG}. For instance, for $M_H=160$ GeV, we obtain with
our  procedure a total $p \bar p  \to H+X$ cross section of  $\sigma^{\rm
tot}=374$ fb, compared to  the value $\sigma^{\rm tot}=439$ fb quoted in 
Ref.~\cite{Tevatron,ggH-FG}. The difference is mainly due to the fact that we
are working in the NNLO approximation in QCD rather than in the NNLL
approximation. As already, mentioned and in accord with Ref.~\cite{ggH-ADGSW},
we believe that only the NNLO result should be considered as the production
cross sections that are used experimentally include only NNLO effects (not to
mention the fact that the $K$--factors for the cross sections with cuts are
significantly smaller than the $K$--factors affecting the total inclusive cross
section, as will be discussed in the next section). A small difference comes
also from the different treatment of the electroweak radiative corrections
(partial factorisation plus mixed QCD--electroweak contributions in our case
versus complete factorisation in Ref.~\cite{ggH-FG}) and another one percent
discrepancy can be attributed to the numerical uncertainties  in the various
integrations of the partonic sections\footnote{We have explicitly verified,
using the program {\tt HRESUM} \cite{HNNLO} which led to the results of
Ref.~\cite{ggH-FG}, that our NNLO cross section is in excellent agreement with
those available in the literature. In particular, for $M_H=160$  GeV and scales
$\mu_R=\mu_F= M_H$, one obtains $\sigma^{\rm NNLO}=380$ fb with  {\tt HRESUM}
compared to $\sigma^{\rm NNLO}=374$ fb in our case; the 1.5\% discrepancy being
due to the different treatment of the electroweak corrections and the
integration errors. Furthermore, setting the renormalisation and factorisation
scales to $\mu_R=\mu_F=\frac12 M_H$, we find $\sigma^{\rm NNLO}=427$ fb which
is in excellent agreement with the value $\sigma^{\rm NNLO}=434$ fb obtained 
in Ref.~\cite{ggH-radja} and with {\tt HRESUM}, as well as the value in the
NNLL approximation when the scales are set at their central values
$\mu_R=\mu_F= M_H$. This gives us confidence that our implementation of the
NNLO contributions in the NLO code {\tt HIGLU},  including the scale
dependence, is correct.}. 

We should also note that for the Higgs mass value $M_H=160$ GeV, we obtain $K
\simeq 2.15$ for the QCD $K$--factor at NLO and $K \simeq 2.8$ at NNLO. These
numbers are slightly different from those presented in Ref.~\cite{ggH-ADGSW},
$K \simeq 2.4$ and $K \simeq 3.3$, respectively. The reason is that the
$b$--quark loop contribution, for which the $K$--factor at NLO is
significantly  smaller than the one for the top quark contribution \cite{SDGZ}
has been ignored for simplicity in the latter paper;  this difference will be
discussed in section 3.2.

In the case of Higgs--strahlung from  $W$ and $Z$ bosons, the central values of
the cross sections that we obtain are comparable to those given in 
Ref.~\cite{Tevatron,HV-all}, with at most a $\sim 2\%$ decrease in the low
Higgs  mass range, $M_H \lsim 140$ GeV. The reason is that the quark and
antiquark densities, which are the most relevant in these processes and are
more under control than the gluon densities, are approximately the same in the
new MSTW2008 and old  MRST2002 sets of PDFs (although the updated set includes
a new fit to run II Tevatron and HERA inclusive jet data). We should note that
for $M_{H}=115$ GeV for which the production cross sections are the largest,
$\sigma^{\rm WH}= 175$ fb and $\sigma^{\rm ZH}=104$ fb, the QCD $K$--factors are
$\sim 1.2 \; (1.3)$ at NLO (NNLO), while the electroweak corrections decrease
the LO cross sections by $\approx -5$\%. The correcting factors do not change
significantly for increasing $M_H$ values for the Higgs mass range relevant at
the Tevatron. 

Finally, the cross sections for the vector boson fusion channel in which the
recent MSTW set of PDFs is used agree well with those given in
Refs.~\cite{Tevatron,TeV4LHC}. In the case of the $t\bar t H$ associated
production process, a small difference is observed compared to
Ref.~\cite{Review} in which the 2005 $m_t=178$ GeV value is used: we have a
few percent increase of the rate due the presently smaller $m_t$ value which
provides more phase space for the process, overcompensating the decrease due to
the smaller top--quark Yukawa coupling.

Before closing this section, let us make a few remarks on the Higgs decay 
branching ratios and on the rates for the various individual channels that are
used to detect the Higgs signal at the Tevatron.  For the the Higgs decays, one
should use  the latest version (3.51) of the program {\tt HDECAY} 
\cite{HDECAY} in which the important radiative corrections to the $H \to WW$
decays \cite{HD-HWW} have been recently implemented. Choosing the option which
allows for the Higgs decays into double off--shell gauge bosons, $H \to V^*
V^*$, which provides the best approximation\footnote{ The options in {\tt
HDECAY} where one or two vector bosons are allowed to be on mass--shell do not
give precise results. In addition, in earlier versions,  there was  an
interpolation which smoothened the transition from below to above the
kinematical threshold, $M_H \approx 2M_W$, i.e. right in the most interesting
Higgs mass region at the Tevatron. The option of both gauge bosons being off
mass--shell should be therefore used.} and using the updated input  parameters
$\alpha_s (M_Z)=0.1172$, $m_t=173.1$ GeV and $m_b^{\rm pole}=4.6$ GeV, one
obtains the results shown in Table 1 for the three dominant decay channels in
the mass range relevant at the Tevatron, $H \to W^*W^*, b \bar b$ and $\tau^+
\tau^-$. These results are slightly different from those given in 
Ref.~\cite{Tevatron}. In particular, the $H \to W^*W^*$ rate that we obtain is
a few percent larger for Higgs masses below $\sim$ 170 GeV.

\TABLE[!h]{\small%
\let\lbr\{\def\{{\char'173}%
\let\rbr\}\def\}{\char'175}%
\renewcommand{\arraystretch}{1.77}
\begin{tabular}{|c||c|c|c|}\hline 
$M_H$ (GeV) & $BR(H\!\to\!W^*W^*)$ & $BR(H\!\to\!b\bar b)$ & $BR(H \! 
\to\!\tau^+ \tau^-)$ %& $BR(H\!\to\!ZZ)$          
\\ \hline
115  & 8.311  & 73.02 & 7.328      \\  \hline 
120  & 13.72  & 67.53 & 6.832      \\  \hline 
125  & 20.91  & 60.44 & 6.161      \\  \hline 
130  & 29.63  & 52.02 & 5.342      \\  \hline 
135  & 39.35  & 42.83 & 4.429     \\  \hline 
140  & 49.45  & 33.56 & 3.493      \\  \hline 
145  & 59.43  & 24.81 & 2.599      \\  \hline 
150  & 69.17  & 16.94 & 1.785      \\  \hline 
155  & 79.11  & 10.60 & 1.060      \\  \hline 
160  & 90.56  & 3.786 & 0.404      \\  \hline 
165  & 95.94  & 1,303 & 0.140      \\  \hline 
170  & 96.41  & 0.863 & 0.093      \\  \hline 
175  & 95.82  & 0.669 & 0.072      \\  \hline 
180  & 93.26  & 0.540 & 0.058      \\  \hline 
185  & 84.51  & 0.419 & 0.046      \\  \hline 
190  & 78.71  & 0.343 & 0.038      \\  \hline 
195  & 75.89  & 0.294 & 0.033      \\  \hline 
200  & 74.26  & 0.259 & 0.029      \\  \hline 
\end{tabular} 
\caption{The branching ratios (in \%) of the main decay channels  of the 
Standard Model Higgs boson using the latest version of the program 
\protect{\tt HDECAY} \cite{HDECAY}.}}

In the interesting range 160 GeV $\le M_H \le 170$ for which the Tevatron 
experiments are most sensitive, one sees that the branching ratio for the $H
\to WW$ is largely dominant, being above 90\%. In addition, in this mass range,
the $gg \to H$ cross section is one order of magnitude larger than the cross
sections for the  $q \bar q \to WH,ZH$ and $qq \to qqH$ processes as for  $M_H
\sim 160$ GeV for instance, one has $\sigma( gg\to H)=374$ fb compared to
$\sigma(WH) \simeq 50$ fb, $\sigma(ZH) \simeq 30$ fb and $\sigma(qqH) \simeq
40$ fb.  Thus, the channel $gg \to H \to W^*W^* $ represents, even before
selection cuts are applied,  the bulk of the events leading to $\ell \ell \nu
\nu +X$ final states, where here $X$ stands for additional jets or leptons
coming from $W,Z$ decays as well as for jets due to the  higher order
corrections to the $gg\to H$ process.  In the lower Higgs mass range, $M_H
\lsim 150$ GeV, all the production channels above,  with the exception of the
vector boson  $qq \to qqH$ channel which can be selected using specific
kinematical cuts, should be taken into account but with the process $q\bar q
\to WH \to \ell \nu b\bar b$ being dominant for $M_H \lsim 130$ GeV. This
justifies the fact that we concentrate on the gluon--gluon fusion and
Higgs--strahlung production channels in this paper.

\section{Theoretical uncertainties in gluon--gluon fusion}

\subsection{The scale uncertainty and higher order effects} 

It has become customary to estimate the effects of the unknown (yet
uncalculated) higher order contributions to production cross sections and
distributions at hadron colliders by studying the variation of these
observables, evaluated at the highest known perturbative order, with the
renormalisation scale $\mu_R$ which defines the strong coupling constant
$\alpha_s$ and the factorisation scale $\mu_F$ at which one performs the
matching between the perturbative calculation of the matrix elements and the
non--perturbative part which resides in the parton distribution functions. The
dependence of the cross sections and distributions on these two scales is in
principle unphysical: when all orders of the perturbative series are summed,
the observables should be scale independent. This scale dependence appears
because the perturbative series are truncated, as only its few first orders are
evaluated in practice, and can thus serve as a guess of the impact of the
higher order contributions.

Starting from a median scale $\mu_0$ which, with an educated guess, is
considered as the most ``natural" scale of the process and absorbs potentially
large logarithmic corrections, the current convention is  to vary these two
scales within the range 
\beq
\mu_0/\kappa \le \mu_R, \mu_F \le \kappa \mu_0 \, .  
\label{scalek}
\eeq
with the constant factor $\kappa$ to be determined. One then uses the following
equations to calculate the deviation of, for instance, a cross section
$\sigma(\mu_R, \mu_F)$  from the central value  evaluated at scales
$\mu_R=\mu_F=\mu_0$,  
\beq
\Delta\sigma_{\mu}^{+} &= &  
\max_{(\mu_{R},\mu_{F})} \sigma(\mu_{R},\mu_{F}) -\sigma(\mu_{R}=\mu_{F}=\mu_0) 
\, , \nonumber \\
\Delta\sigma_{\mu}^{-} &=& 
\sigma(\mu_{R}=\mu_{F}=\mu_0)-\min_{(\mu_{R},\mu_{F})} \sigma(\mu_{R},\mu_{F}) 
\, .  
\label{eqscaleminus}
\eeq
This procedure is by no means a true measure of the higher order effects  and
should be viewed only as providing a guess of the lower limit on the scale
uncertainty. The variation of the scales in the range of eq.~(\ref{scalek}) can
be individual with $\mu_R$ and $\mu_F$ varying independently in this domain,
with possibly some constraints such as $ 1/\kappa \le  \mu_R/\mu_F \le \kappa$
in order not to generate ``artificially large logarithms", or collective when,
for instance,  keeping one of the two scales fixed, say to $\mu_0$,
and vary the  other scale in the chosen domain.  Another possibility which is
often adopted, is to equate the two scales, $\mu_0/\kappa \le \mu_R= \mu_F \le
\kappa \mu_0$,  a procedure that is  possibly more consistent as most PDF sets
are determined and evolved according to $\mu_R=\mu_F$, but which has  no
theoretical ground as the two scales enter different parts of the calculation
(renormalisation versus factorisation). 

In addition, there is a freedom in the choice of the  variation domain for a
given process and, hence, of the constant  factor $\kappa$. This choice  is
again rather subjective: depending on whether one is  optimistic or
pessimistic, i.e. believes or not that the higher order corrections to the 
process are under control, it can range from $\kappa\!=\!2$ to much higher
values.

In most recent analyses of production cross sections at hadron colliders, a
kind of consensus has emerged and the domain,   \beq  \frac12 \mu_0 \le \mu_R,
\mu_F \le 2 \mu_0 \ ,  \ \     \frac12 \le \mu_R/\mu_F \le 2 \, , 
\label{scale2}  \eeq  has been generally adopted for the scale variation. A
first remark is that  the condition $ \frac12 \le \mu_R/\mu_F \le 2$ to avoid
the appearance of large logarithms might seem too restrictive: after all,
these possible large logarithms can be viewed as nothing else than the
logarithms involving the scales and if they are large, it is simply  a
reflection of a large scale dependence. A second remark is that in the case of
processes in which the calculated higher order contributions are small to
moderate and the perturbative series appears to be well behaved\footnote{This
is indeed the case for some important production processes at the Tevatron,
such as  the Drell--Yan process $p\bar p \to V$  \cite{DYNNLO,pp-V}, weak boson
pair production   \cite{pp-VV} and even top quark pair production \cite{pp-tt}
once the central scale is taken  to be $\mu_0=m_t$, which have moderate QCD
corrections.}, the choice of such a
narrow domain for the scale variation with  $\kappa=2$, appears  reasonable.
This, however, might not be true in processes in which the calculated radiative
corrections turn out to be extremely large. As the higher order contributions
might also be significant in this case, the variation domain of the
renormalisation and factorisation scales should be extended and a range with a
factor $\kappa$ substantially larger than two seems more
appropriate\footnote{This would have been the case, for instance, in top--quark
pair production at the Tevatron if the central scale were fixed to the more
''natural" value $\mu_0=2m_t$ (instead of the value $\mu_0={m_t}$ usually taken
\cite{pp-tt}) and a scale variation within  $\frac14 M_H \le \mu_R,\mu_F \le 4
M_H$ were adopted. Another well known example is Higgs production in
association with $b$--quark pairs in which the cross section can be determined
by evaluating the  mechanism $gg/q\bar q  \to b\bar b H$ \cite{pp-bbH1} or  
$b\bar b$ annihilation, $b\bar b\to H$ \cite{pp-bbH2}. The two calculations 
performed at NLO for the former process and NNLO for the later one, are
consistent only if the central scale is taken to be $\mu_0 \approx \frac14 M_H$
instead of the more ''natural" value $\mu_0 \approx M_H$ \cite{pp-bbH3}. Again,
without prior knowledge of the higher order corrections,  it would have been
wiser, if the central scale $\mu_0=M_H$ had been  adopted, to assume a wide domain,
e.g.  $\frac14 M_H \le \mu_R,\mu_F \le 4 M_H$,  for the scale variation. Note
that even for the scale choice  $\mu_0 \approx \frac14 M_H$, the $K$--factor
for the $gg \to b\bar b H$ process remains very large, $K_{\rm NLO} \approx 2$
at the Tevatron. In addition, here, it is the factorisation scale $\mu_F$ which
generates the large contributions $\propto \ln(\mu_F^2/m_b^2)$ and not the
renormalisation scale which can be thus kept at the initial value $\mu_R 
\approx M_H$.}. 

In the case of the $gg \to H$ production process, the most natural value for
the median scale is the Higgs mass itself, $\mu_0=M_H$, and the effects of the
higher order contributions to the cross section is again usually estimated by
varying $\mu_R$ and $\mu_F$ as in eq.~(\ref{scale2}), i.e. with the choice
$\frac1\kappa \le \mu_R/\mu_F\le \kappa$ and $\kappa=2$. At the Tevatron, one
obtains a variation  of approximately $\pm 15\%$  of the NNLO cross section
with this specific choice \cite{ggH-NNLO1,ggH-NNLO2} and the uncertainty drops
to the level of $\approx \pm 10\%$ in the NNLL approximation. Note that in some
analyses, see e.g. Ref.~\cite{ggH-radja}, the central scale $\mu_0=\frac12 M_H$
is chosen for the NNLO cross section to mimic the soft--gluon resumation  at
NNLL \cite{ggH-resum}, and the variation domain  $\frac14 M_H \le \mu_R=\mu_F
\le  M_H$ is then adopted, leading also to a $\approx 15\%$ uncertainty  

Nevertheless, as the $K$--factor is  extraordinarily large in the $gg \to H$
process, $K_{\rm NNLO} \approx 3$, the domain  of eq.~(\ref{scale2}) for  the
scale variation seems too narrow. If this scale domain was chosen for the LO
cross section for instance, the maximal value of $\sigma(gg \to H)$ at LO would
have never caught, and by far, the value of $\sigma(gg \to H)$ at NNLO,  as it
should be the case if the uncertainty band with $\kappa=2$ were indeed the
correct  ``measure" of the higher order effects. Only for a much larger value
of $\kappa$ that this would have been the case. 

Here, we will use a criterion which allows an empirical evaluation of the
effects of the still unknown high orders of the perturbative series and, hence,
the choice of the variation domain of the factorisation and renormalisation
scales in a production cross section (or distribution). This is done in two
steps:

$i)$  The domain of scale variation, $\mu_0/{\kappa} \le \mu_R, \mu_F \le
\kappa \mu_0$, is derived by calculating the factor $\kappa$ which allows the
uncertainty band of the lower order cross section resulting from the
variation of  $\mu_R$ and $\mu_F$, to reach the central value (i.e. with
$\mu_R$ and $\mu_F$ set to $\mu_0$), of the cross section  that has been
obtained at the higher perturbative order.

$ii)$ The scale uncertainty on the cross section  at the higher perturbative
order is then taken to be the band obtained for a variation of the scales
$\mu_R$ and $\mu_F$ within the same range and, hence, using the same $\kappa$
value.  

In the case of the $gg \to H$ process at the Tevatron, if the lower order 
cross section is taken to be simply $\sigma^{\rm LO}$ and the higher order one 
$\sigma^{\rm NNLO}$, this is exemplified in the left--hand side of Fig.~2. The
figure shows the uncertainty band  of $\sigma^{\rm LO}$ resulting from a scale
variation in the domain $M_H/ {\kappa} \le \mu_R,\mu_F \le \kappa M_H$  with
$\kappa=2,3,4,5$, which is then compared to $\sigma^{\rm NNLO}$  evaluated at
the central scale $\mu_R=\mu_F= M_H$.  One first observes that, as expected,
the uncertainty bands are larger with increasing values of $\kappa$.  

The important observation that one can draw from this figure is that it is only for
$\kappa\!=\!5$, i.e. a variation of the scales in a range that is much wider
than the one given in eq.~(\ref{scale2})  that the uncertainty band of the LO
cross section becomes very close to (and still does not yet reach for low  Higgs
mass values) the curve giving the NNLO result. Thus, as the scale uncertainty
band of  $\sigma^{\rm LO} (gg \to H)$ is supposed to provide an estimate of
the resulting cross section at NNLO and beyond, the range within which the two
scales $\mu_R$ and $\mu_F$ should be varied must be significantly larger than
$\frac12 M_H \le \mu_R, \mu_F \le 2 M_H$. On should not impose a restriction on
$\mu_R/\mu_F$ and consider at least the range\footnote{
Note that, in this case, the maximal LO cross section is obtained
for small values of the two scales $\mu_R$ and $\mu_F$. In fact, if the central
value for the scales had been chosen to be $\mu_F=\mu_R= \frac15 M_H$ for
instance, one would have obtained  at LO, NLO and NNLO a cross section
$\sigma^{\rm LO}=360$ fb, $\sigma^{\rm  NLO}=526$ fb and $\sigma^{\rm NNLO}
=475$ fb for the Higgs mass value $M_H=160$ GeV. The increase of the LO cross
section by a factor of $\approx 2.8$, compared to the case $\mu_F=\mu_R=M_H$
where one has $\sigma^{\rm LO}=129$ fb for the chosen $M_H$ value, has absorbed
the bulk of the higher order corrections. This allows a good convergence of the
perturbative series as in this case one has $K_{\rm NLO}=1.46$ and  $K_{\rm
NNLO}= 1.32$, which seems to stabilize the cross section between the NLO and
NNLO values. This nice picture is not spoilt by soft--gluon resumation which
leads for such a scale to $\sigma^{\rm NNLL}=459$ fb and, hence, the
$K$--factor turns  to $K_{\rm NNLL}=1.28$ which is only a few percent lower than
$K_{\rm NNLO}$. Thus, it might have been worth to choose $\mu_0=\frac15 M_H$ as
the central scale from the very beginning, although this particular value does
not look very ``natural" a priori. We also point out the fact that the choice
$\mu_0= \frac15 M_H$ for the central scale, provides  an example of a reduction
of the cross section when higher order contributions are taken into account as
$K_{\rm NNLL} <K_{\rm NNLO} <K_{\rm NLO}$.} $\frac15 M_H \le \mu_R, \mu_F \le
5M_H$.

\begin{figure}[!h]
\vspace*{5mm}
\begin{center}
\mbox{
\hspace*{-.5cm}
\epsfig{file=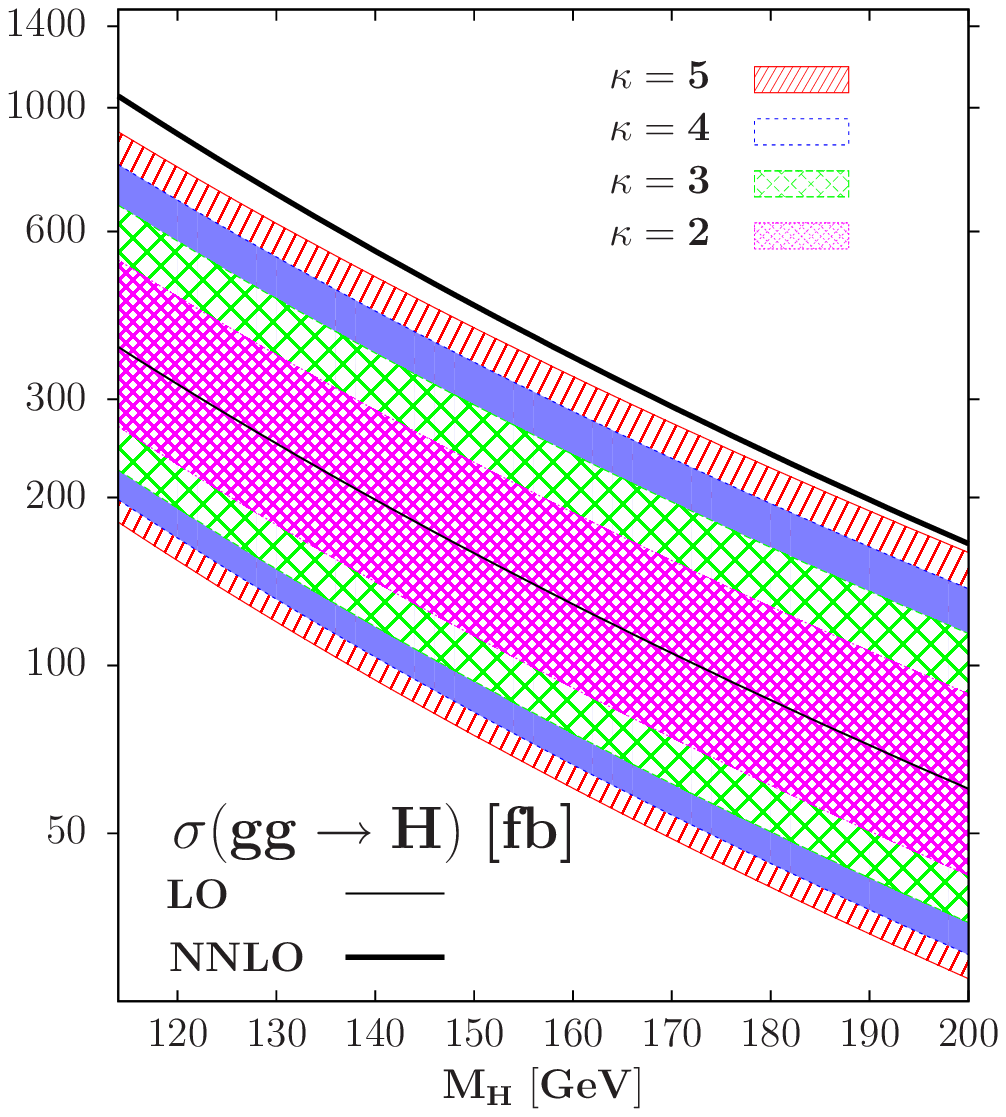,width=7.7cm} \hspace*{-0.1cm}
\epsfig{file=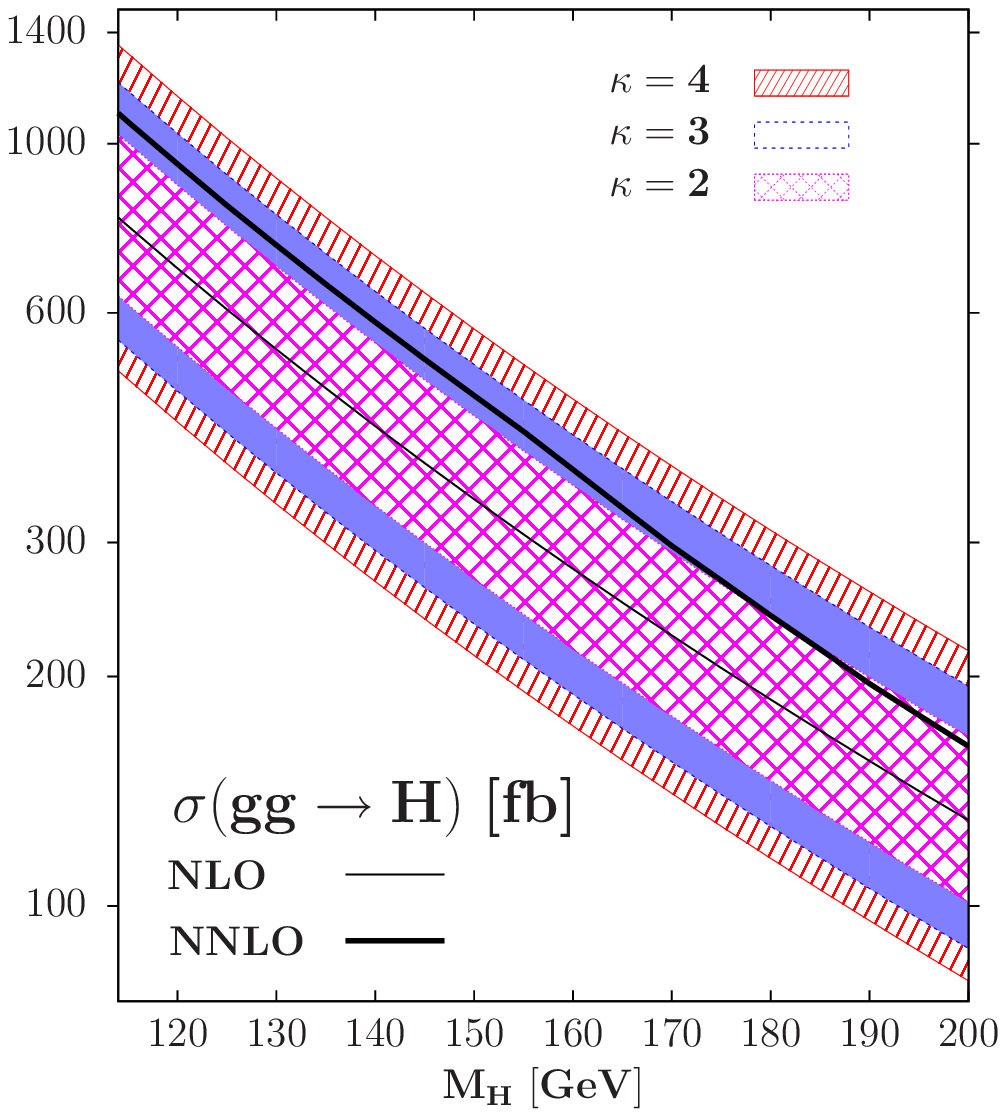,width=7.7cm}
}
\end{center}
\vspace*{-4mm}
\caption[]{Left: the scale dependence of $\sigma^{\rm LO} (gg \to H)$ at the 
Tevatron as a function of $M_H$ for scale variations $M_H/{\kappa} \le \mu_R, 
\mu_F \le \kappa M_H$ with $\kappa=2,3,4$ and 5 compared to $\sigma^{\rm NNLO}$ 
for the central  scale choice $\mu_R=\mu_F=M_H$. Right: the scale dependence of 
$\sigma^{\rm NLO} (gg \to H)$ at the Tevatron as a function of $M_H$ for 
variations $M_H/{\kappa} \le \mu_R, \mu_F \le \kappa M_H$ with $\kappa=2,3$ 
and 4 compared to  $\sigma^{\rm NNLO}$ evaluated at the central scale $\mu_R=
\mu_F=M_H$.}
%\vspace*{-3mm}
\end{figure}

Nevertheless, one might be rightfully reluctant to use $\sigma^{\rm LO}$  as a
starting point for estimating the higher order effects, as it is well known
that it is only after including at least the next--order QCD corrections that a
cross section is somewhat stabilized and, in the particular case of the $gg\to
H$ process, the LO cross section does not describe correctly the kinematics as,
for instance, the Higgs transverse momentum is zero at this order. We thus
explore also the scale variation of the  NLO cross section $\sigma^{\rm NLO}$ 
instead of that of $\sigma^{\rm LO}$ and compare the resulting uncertainty band
to the central value of the cross  section again at NNLO (we refrain here from
adding the $\sim 15\%$ contribution at NNLL as well as those arising from
higher order corrections, such as the estimated ${\rm N^3LO}$ correction
\cite{ggH-N3LO}).  

The scale uncertainty bands of $\sigma^{\rm NLO}$ are shown in the right--hand
side of Fig.~2 as a function of $M_H$ again for scale variation in the domain
$M_H/{\kappa} \le \mu_R,\mu_F \le \kappa M_H$  with $\kappa=2,3$ and 4,  and
are compared to $\sigma^{\rm NNLO}$  evaluated at the central scale
$\mu_R=\mu_F= M_H$. One can see that, in this case, the uncertainty band for
$\sigma ^{\rm NLO}$ shortly falls to reach  $\sigma^{\rm NNLO}$ for
$\kappa\!=\!2$ and only for $\kappa\!=\!3$ that this indeed occurs in the
entire $M_H$ range. 

Thus, to attain the NNLO values of the $gg \to H$ cross section  at the
Tevatron with the scale variation of the NLO cross section, when both cross
sections are taken at the central scale choice\footnote{We note that one could
choose the central scale value $\mu_0=\frac12 M_H$  \cite{ggH-radja}, instead
of $\mu_0=M_H$, which seems to better describe the essential features of the
kinematics of the process, and in this case, a variation within a factor of two
from this central value would have been  sufficient for $\sigma^{\rm NLO}$ to
attain $\sigma^{\rm NNLO}$. We thank Babis Anastasiou for a discussion on this
point.} $\mu_R=\mu_F=\mu_0=M_H$, one needs to chose the values $\kappa=3$, and
hence a domain of scale variation that is wider than that given in
eq.~(\ref{scale2}). This choice of the domains of scale variation might seem
somewhat conservative at first sight.  However, we  emphasise again that in
view of the  huge QCD corrections which affect the cross section of this
particular process, and which almost jeopardize the convergence of the
perturbative series, this choice appears to be justified. In fact, this scale
choice is not so unusual and in  
Refs.~\cite{ggH-NNLO2,ggH-NNLO3,ggH-resum,largekappa} for instance, scale
variation domains comparable to those discussed here, and sometimes even wider,
have been used for illustration.

Thus, in our analysis, rather than taking the usual choice for the scale domain
of variation with $\kappa=2$ given in  eq.~(\ref{scale2}), we will adopt the
slightly more conservative possibility given by the wider variation 
domain\footnote{One might argue that since in the case of $\sigma (gg \to H)$,
the  NLO and NNLO contributions are both positive and increase the LO rate, one
should expect a positive contribution from higher orders (as is the case for
the re-summed NNLL contribution) and, thus, varying the scales using
$\kappa=2$  is more conservative, as the obtained maximal value of the cross
section  would be smaller than the value that one would obtain for e.g.
$\kappa=3$. However, one should not assume that the higher order contributions
always increase the lower order cross sections.  Indeed, as already mentioned,
had we taken the central scales at $\mu_R=\mu_F=\frac15 M_H$, the NNLO (and
even NNLL) corrections would have  reduced the total cross section evaluated at
NLO. Hence, the higher order contributions to $\sigma(gg\to H)$ could well be
negative beyond NNLO and could bring the value of the production cross section
close to the lower range of the scale uncertainty band of $\sigma^{\rm NNLO}$.
Another good counter-example of a cross section that is reduced by the higher
order contributions is the process of  associated Higgs production with top
quark pairs at the Tevatron where the NLO QCD corrections decrease the LO cross
section by $\sim 20\%$  \cite{ttH-NLO} once the central scale is chosen to be
$\mu_0=\frac12(2m_t+M_H)$.}
\beq     
\frac13 M_H \le \mu_R, \mu_F \le 3M_H\, .  \label{scale3}  
\eeq 

Having made this choice for the factor $\kappa$, one can turn to the
estimate of the higher order effects of $\sigma ({ gg \to H})$ evaluated at
the highest perturbative order that we take  to be NNLO, ignoring again the 
known small contributions beyond this fixed order.  

The uncertainty bands resulting from scale variation  of  $\sigma^{\rm NNLO}(gg
\to H)$  at NNLO in the domains given by eqs.~(\ref{scale2}) and (\ref{scale3})
are shown in Fig.~3 as a function of $M_H$. As expected, the scale  uncertainty
is  slightly larger for $\kappa=3$ than for $\kappa=2$.  For instance,
for $M_H=160$ GeV, the NNLO cross section varies by up to $\sim \pm {21}\%$ from
its central value, $\sigma^{\rm NNLO}=374\pm 80~{\rm fb}$, compared to the
$\approx \pm  {14}\%$ variation that one obtains for $\kappa=2$, $\sigma^{\rm
NNLO}=374\pm 52~{\rm fb}$. The minimal cross section is obtained for the
largest  values of the two scales, $\mu_F=\mu_R= \kappa M_H$, while the maximal
value is obtained for  the lowest value of the renormalisation scale,
$\mu_R=\frac1\kappa M_H$, almost independently of the factorisation scale
$\mu_F$, but with a slight preference for the  lowest $\mu_F$ values,
$\mu_F=\frac1\kappa M_H $. 

\begin{figure}[!h]
%\vspace*{.2cm}
\begin{center}
\hspace*{-1.cm}
\epsfig{file=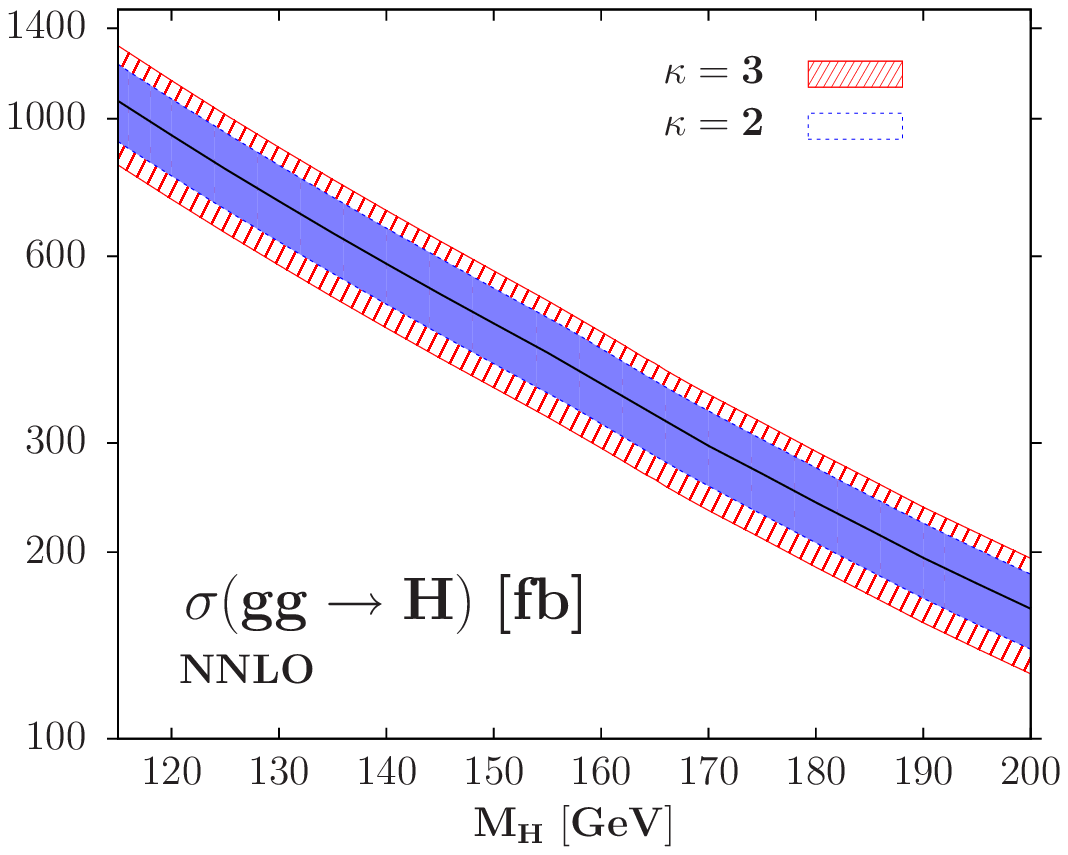,width=10.cm} 
\end{center}
\vspace*{-5mm}
\caption[]{ 
The uncertainty bands of the NNLO $gg \to H$ cross section  at the Tevatron as 
a function of $M_H$ for scale variation in the domains $\frac13 M_H \le \mu_R, 
\mu_F \le 3 M_H$ and $\frac12 M_H \le  \mu_R, \mu_F \le 2 M_H$.}
\vspace*{-3mm}
\end{figure}

We should note that the $\approx 10\%$ scale uncertainty obtained in 
Ref.~\cite{ggH-FG} and adopted by the CDF/D0 collaborations \cite{Tevatron} is
even smaller than the ones discussed above. The reason is that it is the
resumed NNLL cross section, again with $\kappa\!=\!2$ and $\frac12\! \le\!
\mu_R/\mu_F\! \le\! 2$, that was considered, and the scale variation of
$\sigma^{\rm NNLL}$ is reduced compared to that of  $\sigma^{\rm NNLO}$ in this
case. As one might wonder if this milder dependence also occurs for our adopted
$\kappa$ value, we have explored the scale variation of  $\sigma^{\rm NNLL}$ in
the case of  $\kappa=3$, without the restriction $\frac13\! \le\!
\mu_R/\mu_F\!  \le\! 3$. Using again the program {\tt HRESUM} \cite{HNNLO}, we
find that the difference between the maximal value of the NNLL cross section,
obtained for $\mu_R \approx M_H$ and $\mu_F \approx 3M_H$, and its minimal
value, obtained for $\mu_F \approx  \frac13 M_H$ and $\mu_R \approx 3M_H$, is
as large as in the NNLO case (this is also true for larger $\kappa$ values).
The maximal decrease and maximal increase of $\sigma^{\rm NNLL}$ from the
central value are still of about $\pm 20\%$ in this case. Hence, the relative
stability of the NNLL cross section against scale variation, compared to the
NNLO case,  occurs only  for $\kappa=2$  and may appear as accidentally due to 
a restrictive choice of the variation domain. However, if the additional
constraint $1/\kappa \le \mu_F/\mu_R \le \kappa$ is implemented, the situation
would  improve in the NNLL case, as  the possibility $\mu_F \approx 
\frac1\kappa M_H$ and $ \mu_R \approx \kappa M_H$ which minimizes  $\sigma^{\rm
NNLL}$  would be absent and the scale variation reduced. Nevertheless, even in
this case, the variation  of $\sigma^{\rm  NNLL}$ for $\kappa=3$ is of the
order of $\approx \pm 15\%$ and, hence, the scale uncertainty is  larger than 
what is obtained in  the domain of eq.~(\ref{scale2}). 

Finally, another reason for a more conservative choice of the scale variation 
domain for $\sigma^{\rm NNLO}$, beyond the minimal $\frac 12 M_H \le
\mu_R,\mu_F \le 2 M_H$ range,  is that it is well known that the QCD
corrections are significantly larger for the total inclusive cross section than
for that on which basic selection cuts  are applied; see e.g.
Ref.~\cite{ggH-cuts}.  This can be seen  from the recent analysis of 
Ref.~\cite{ggH-ADGSW}, in which  the higher order corrections to the inclusive
cross section for the  main Tevatron Higgs signal, $gg \to H \to \ell \ell \nu
\nu$, have been compared  to those affecting the cross section when selection
cuts, that are very similar to those adopted by the CDF and D0 collaborations 
in their analysis (namely lepton selection and isolation, a minimum
requirement for the missing transverse energy due to the neutrinos, and a veto
on hard jets to suppress the $t\bar t$ background), are applied.  The output of
this study is that the $K$--factor
for the cross section after cuts  is $\sim 20$--30\% smaller than the
$K$--factor for the inclusive total cross section (albeit with a reduced scale
dependence).  For instance,  one has $K^{\rm NNLO}_{\rm cuts}=2.6$ and $K^{\rm
NNLO}_{\rm total}= 3.3$ for  $M_H=160$ GeV and scales set to
$\mu_F=\mu_R=M_H$. 

Naively, one would expect that this $\sim 20$--30\% reduction of the higher
order QCD corrections when selection cuts are applied,  if not implemented from
the very beginning in the normalisation of the cross section after cuts that is
actually used by the experiments (which would then reduce the acceptance of the
signal events,  defined as $\sigma^{\rm NNLO}_{\rm cuts}/\sigma^{\rm NNLO}_{\rm
total}$), to be at least  reflected in the scale variation of the inclusive
cross section and, thus, accounted for in the theoretical uncertainty. This
would be partly the case for scale variation within a factor $\kappa=3$ from
the central scale, which leads to a maximal reduction of the $gg \to H \to \ell
\ell \nu \nu$ cross section by about $20\% $, but  not with the choice
$\kappa=2$  made in  Refs.~\cite{Tevatron} which would have led to a possible
reduction of the cross section by $\approx 10$\% only\footnote{The discussion
is, however,  more involved as one has to consider the efficiencies obtained
with the NNLO calculation compared to that obtained with the Monte--Carlo used
by the experiments; see  Ref.~\cite{ggH-ADGSW}.}. 

\subsection{Uncertainties due to the effective approach}

While both the QCD and electroweak radiative corrections to the process $gg \to
H$ have been calculated exactly at NLO, i.e taking into account the finite mass
of the particles running in the loops,  these corrections are derived at NNLO 
only in an effective approach in which the loop particles are assumed  to be
very massive, $m \gg M_H$, and integrated out. At the Born level,  taking into
account only the dominant contribution of the top quark loop and  working in
the limit $m_t \to \infty$ provides an approximation \cite{SDGZ,Michael-rev}
that is only good at the 10\% level for Higgs masses below the $t\bar t$
kinematical threshold, $M_H \lsim 350$ GeV. The difference from the exact
result is mainly due to the absence of the  contribution of the $b$--quark
loop: although the $b$--quark mass is small, the $gg \to H$ amplitude exhibits
a dependence  $\propto m_b^2/M_H^2 \times \log^2 (m_b^2/M_H^2)$ which, for
relatively low values of the Higgs mass,  generates a non--negligible 
contribution that interferes destructively with the dominant top--quark loop
contribution. In turn, when considering only the top quark loop in the $Hgg$
amplitude, the approximation $m_t  \to \infty$ is extremely good for Higgs
masses below $2m_t$, compared to the amplitude with the exact top quark mass
dependence.

In the NLO approximation for the QCD radiative corrections, it has been shown
\cite{SDGZ} that the exact $K$--factor when the full dependence on the top and
bottom quark masses is taken into account, $K^{\rm exact}_{\rm NLO}$, is
smaller than the $K$ factor obtained in the approximation in which only the top
quark contribution is included  and the asymptotic limit $m_t \to \infty$ is
taken, $K^{m_t\!\to\!\infty}_{\rm NLO}$. The reason is that when only the
$b$--quark loop contribution is considered in the $Hgg$ amplitude (as in the
case of supersymmetric theories in which the $b$--quark Yukawa  coupling is
strongly enhanced compared to its Standard Model value \cite{Review2}), the
$K$--factor for the $gg \to H$ cross section at the Tevatron is about $K
\sim 1.2$ to  1.5, instead of $K \sim 2.4$ when only the top quark is included
in the loop. The approximation of infinite loop particle mass significantly
improves when the full $t,b$ mass dependence is included in the LO order cross
section and $\sigma_{\rm NLO}^{m_t\! \to \! \infty} =  K^{m_t\!\to\!
\infty}_{\rm NLO} \times \sigma_{\rm LO} (m_t,m_b)$ gets closer to the cross
section $\sigma^{\rm exact}_{\rm NLO}$ in which the exact $m_t, m_b$ dependence
is taken into account. In fact, this approximation works at the 10\% level 
even beyond the  $M_H \gsim 2m_t$ threshold where the $Hgg$ amplitude develops
imaginary parts that do not appear in the effective approach.

 The difference between $\sigma^{\rm exact}_{\rm NLO}$ and $\sigma^{m_t \! \to
\! \infty}_{\rm NLO}$ at Tevatron energies is shown in Fig.~4 as a function of
the Higgs mass and, as one can see, there is a few percent discrepancy between
the two cross sections. As mentioned previously, in the Higgs mass range 115
GeV$\lsim M_H \lsim 200$ GeV relevant at Tevatron energies, this difference is
solely due to the absence of the $b$--quark loop contribution and its
interference with the top quark loop in the $Hgg$ amplitude and not to the
fact that the limit $m_t \gg M_H$ is taken. 

\begin{figure}[!t]
\begin{center}
\epsfig{file=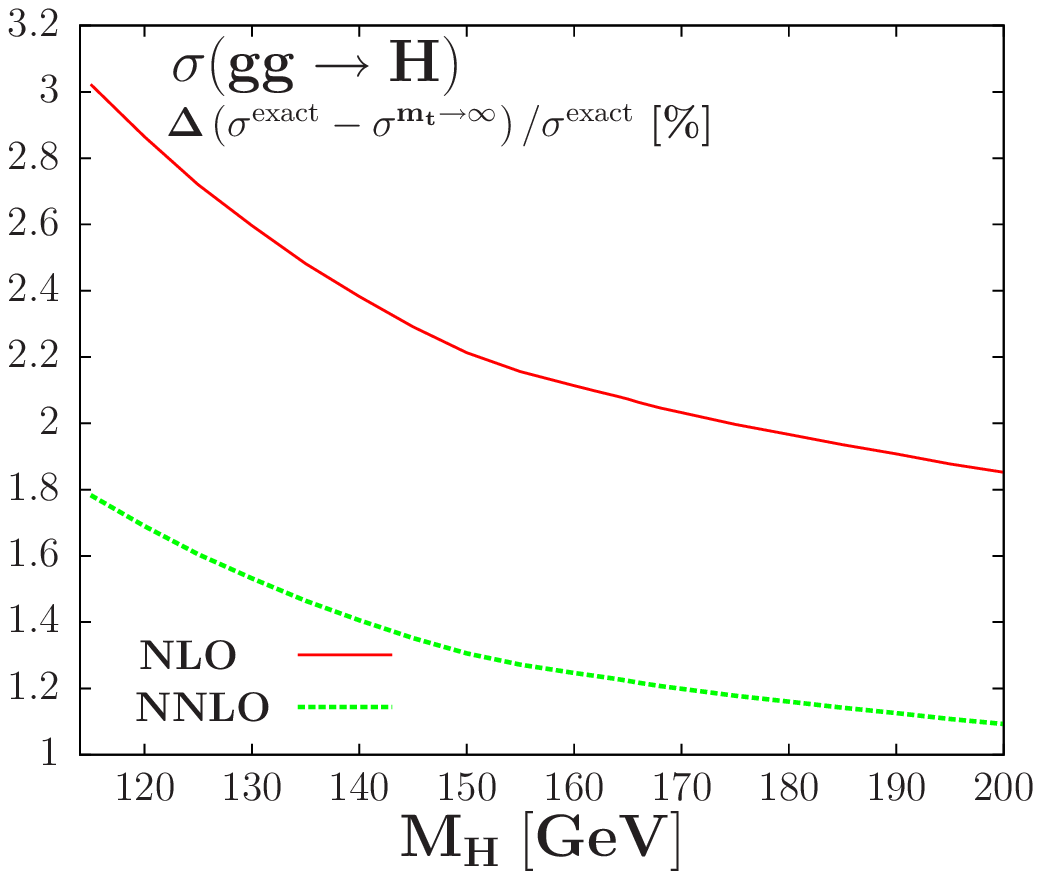,width=9cm}
\end{center}
\vspace*{-5mm}
\caption[]{Relative difference (in \%) at Tevatron energies and as a function 
of $M_H$ between the exact NLO and NNLO $gg\to H$ cross sections $\sigma^{\rm 
exact}_{\rm NLO/NNLO}$ and the cross section in the effective approach with an 
infinite top quark mass $\sigma^{m_t \! \to \! \infty}_{\rm NLO/NNLO}$.}
\vspace*{-2mm}
\end{figure}

At NNLO, because of the complexity of the calculation, only the result in the
effective approach in which the loop particle masses are assumed to be infinite
is available. In the case of the NNLO QCD corrections 
\cite{ggH-NNLO1,ggH-NNLO2,ggH-NNLO3}, the $b$--quark loop contribution and its
interference with the contribution of $t$--quark loop is therefore missing.
Since the NNLO correction increases the cross section by $\sim 30\%$, one might
wonder if this missing piece does not lead to an overestimate of the total
$K$--factor.  We will assume that it might be indeed the case and assign an
error on the NNLO QCD result  which is  approximately the difference between
the exact result $\sigma^{\rm exact}_{\rm NLO}$ and the approximate result
$\sigma^{m_t \! \to \! \infty}_{\rm NLO}$  obtained at NLO and shown in Fig.~4,
but rescaled with the relative magnitude of the $K$--factors that one obtains
at NLO and NNLO, i.e. $K_{\rm NLO}^{m_t \to \infty}/K_{\rm NNLO}^{m_t \to
\infty}$. This leads to an uncertainty on the NNLO cross section which ranges
from $\sim \pm 2\%$  for low Higgs values $M_H  \sim 120$ GeV at which the
$b$--quark loop contribution is significant at LO, to the level of $\sim \pm
1\%$ for Higgs masses above $M_H \sim 180$ GeV for which the $b$--quark loop
contribution is much smaller.

In addition one should assign to the $b$--quark contribution an error
originating from the freedom in choosing the input value of the $b$--quark mass
in the loop amplitude and the scheme in which it is defined\footnote{We thank
Michael Spira for reminding us of this point.}.  Indeed, besides the
difference  obtained when using the $b$--quark pole mass, $M_b^{\rm
pole}\approx 4.7$ GeV, as is done here or the running $\overline{ \rm MS}$
mass evaluated at the scale of the $b$--quark mass,  $\bar m_b^{\rm MS} (M_b)
\sim 4.2$ GeV, there is an additional $\frac43  \frac{\alpha_s}{\pi}$ factor
which enters the cross section when switching from the on--shell to the
$\overline{\rm MS}$ scheme. This leads to an error of approximately 1\% on the
total cross section, over the $M_H$ range that is relevant at the Tevatron.  In
contrast,  according to very recent calculations \cite{ggH-NNLO-mt}, the  $m_t
\to \infty$ limit is a rather good approximation for the top--quark loop
contribution to $\sigma(gg \to H)$ at NNLO as the higher order terms, when
expanding the amplitude in power series of  $M_H^2/(4m_t^2)$, lead to a
difference that is smaller than one percent for  $M_H \lsim  300$ GeV.

We turn now our attention to the electroweak radiative corrections and also
estimate their associated error. As mentioned previously, while the ${\cal
O}(\alpha)$ NLO corrections have been calculated with the exact dependence on
the loop particle masses \cite{ggH-actis}, the mixed QCD--electroweak
corrections  due to light quark loops at ${\cal O}(\alpha \alpha_s)$ have been
evaluated \cite{ggH-radja} in the effective theory approach where the $W,Z$
bosons have been integrated out and which is only valid for $M_H \ll M_W$.
These contributions are approximately equal to the difference between the exact
NLO electroweak corrections when evaluated in the complete factorisation and
partial factorization schemes \cite{ggH-radja}. 

However, as the results for the mixed corrections are  only valid at most for
$M_H < M_W$ and given  the fact that  the companion $\delta_{\rm EW}$
electroweak correction at ${\cal O}(\alpha)$ exhibits a completely different
behavior  below and above the $2M_W$ threshold\footnote{Indeed, the NLO
electroweak correction $\delta_{EW}$  of Ref.~\cite{ggH-actis} is positive
below the $WW$ threshold $M_H \lsim 2M_W$ for which the effective approach is
valid in this case and turns to negative for $M_H \gsim 2M_Z$ for which the
effective approach cannot be applied and the amplitude develops imaginary 
parts. This behavior can also be seen in Fig.~5 which, up to the overall
normalisation, is to a very good approximation  the $\delta_{\rm EW}$
correction factor given in  Fig.~1 of Ref.~\cite{ggH-actis} for $M_H \lsim
2M_Z$.},  one should be cautious and assign an uncertainty to this mixed
QCD--electroweak correction. Conservatively, we have chosen to assign an error
that is of the same size as the  ${\cal O}(\alpha \alpha_s)$ contribution
itself. This is equivalent to assigning an error to the full  ${\cal
O}(\alpha)$ contribution that amounts to the difference between the correction
obtained in the complete factorisation and partial factorisation schemes as
done in Ref.~\cite{ggH-actis}. As pointed out in the latter reference, this 
reduces to adopting the usual and well--established procedure that has been
used at LEP for attributing uncertainties due to unknown higher order effects. 
Doing so, one obtains an uncertainty ranging  from 1.5\% to 3.5\%  for Higgs
masses below $M_H \lsim 2M_W$ and below 1.5\% for larger Higgs masses as is
shown in Fig.~5. 

\begin{figure}[!t]
\begin{center}
\vspace*{3mm}
\epsfig{file=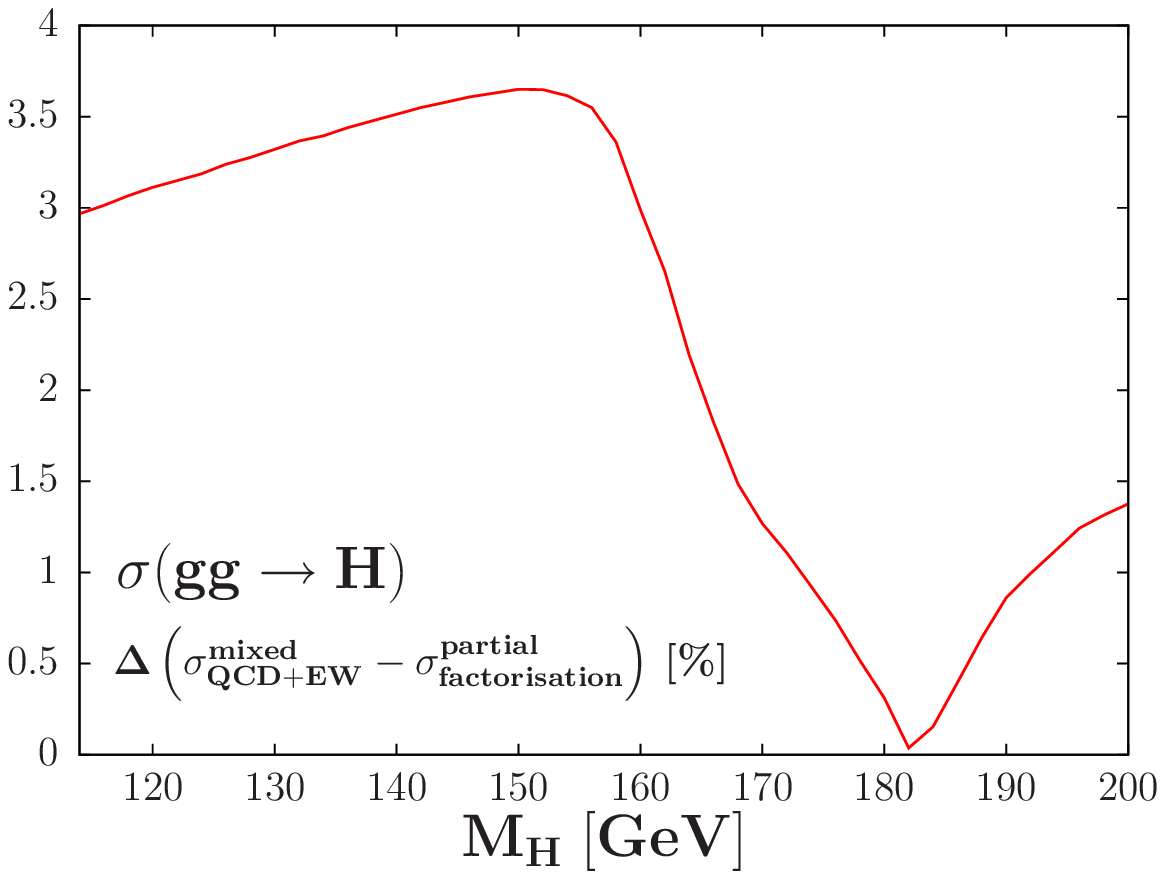,width=9.cm}
\end{center}
\vspace*{-5mm}
\caption[]{Relative difference (in \%) between the complete factorisation and
partial factorisation approaches for the electroweak radiative corrections to
the NLO $gg \to H$ cross section at the Tevatron  as a function of $M_H$.}
\vspace*{-2mm}
\end{figure}

Finally, we should note that we do not address here the issue of the threshold
effects from virtual $W$ and $Z$ bosons which lead to spurious spikes in the
${\cal O}(\alpha)$ electroweak correction in the  mass range $M_H=160$--190 GeV
which includes the Higgs mass domain that is most relevant at the Tevatron (the
same problem occurs in the case of the $p\bar p \to HV$ cross sections once the
electroweak corrections are included). These singularities are smoothened by
including the finite widths of the $W/Z$ bosons, a procedure which might
introduce potential additional theoretical ambiguities that we will ignore in
the present analysis.

\subsection{Uncertainties from the PDFs and $\alpha_s$} 

Another major source of theoretical uncertainties on production cross sections
and distributions at hadron colliders is due to the still imperfect
parametrisation of the parton distribution functions. Within a given
parametrisation, for example the one in the MSTW scheme, these uncertainties
are estimated as follows \cite{PDF-MRST,PDF-as,PDF-Samir}. The scheme is based
on a matrix method which enables a characterization of a parton parametrization
in the neighborhood of the global $\chi^2$ minimum fit and gives an access to
the uncertainty estimation through a set of PDFs that describes this
neighborhood. The corresponding PDFs are constructed by: $i)$ performing  a
global fit of the data using $N_{\rm PDF}$ free parameters ($N_{\rm PDF}=15$ or
20, depending on the scheme); this provides the nominal PDF or reference set
denoted by $S_0$; $(ii)$ the global $\chi^2$ of the fit  is increased to a
given value $\Delta \chi^2$ to obtain the error matrix; $(iii)$ the error
matrix is diagonalized to obtain $N_{\rm PDF}$ eigenvectors corresponding to
$N_{\rm PDF}$ independent directions in the parameter space; $(iv)$ for each
eigenvector, up and down excursions are performed in the tolerance gap,
$T=\sqrt{\Delta \chi^2_{\rm global}}$, leading to $2N_{\rm PDF}$ sets of new
parameters, denoted by $S_i$, with $i=1, 2N_{\rm PDF}$. 

These sets of PDFs  can be used to calculate the uncertainty on a cross section
$\sigma$ in the following way: one first evaluates the cross section with the
nominal PDF $S_0$ to obtain  the central value $\sigma_0$, and then calculates 
the cross section with  the $S_i$ PDFs, giving $2N_{\rm PDF}$ values $\sigma_i$,
and defines, for each $\sigma_i$ value, the deviations 
\beq
\sigma_i^\pm =\mid \sigma_i -\sigma_0\mid \ \ {\rm when} \  
\sigma_i \ ^{>}_{<}  \sigma_0
\eeq
The uncertainties are summed quadratically to calculate the cross section, 
including the error from the PDFs that are given at the 90\% confidence level
(CL),
\beq
\sigma_0|^{+\Delta \sigma^+_{\rm PDF} }_{- \Delta \sigma^-_{\rm PDF}} \ \ 
{\rm with} \ \Delta \sigma^\pm_{\rm PDF}  = \left( \sum_i \sigma_i^{\pm 2} 
\right)^{1/2}
\eeq

The procedure outlined above has been applied to estimate the PDF uncertainties
in the Higgs production cross sections in the gluon--gluon fusion mechanism  at
the Tevatron in Refs.~\cite{ggH-FG,ggH-radja}. This has led to a 90\% CL 
uncertainty of $\approx 6\%$ for the low mass range $M_H \approx 120$ GeV to
$\approx  10\%$ in the high mass range, $M_H \approx 200$ GeV. These
uncertainties have been adopted in the CDF/D0 combined Higgs search and
represent the second largest source of errors after  the scale variation. We
believe that, at least in the case of the $gg$ fusion mechanism, restricting to
the procedure described above largely underestimates the PDF uncertainties for
at least the two reasons discussed below.

First of all, the MSTW collaboration \cite{PDF-MSTW} is not the only one which
uses the above scheme for PDF error estimates, as the CTEQ \cite{PDF-CTEQ} and
ABKM \cite{PDF-Alekhin}  collaborations, for instance, also provide similar
schemes (besides the NNPDF set \cite{PDF-NN},  an additional NNLO PDF set
\cite{PDF-JR} has recently appeared and it also allows for error estimates). It
is thus more appropriate to compare the results given by the three different
sets and take into account the possibly different errors that one obtains. In
addition, as the parameterisations of the PDFs are different in the three
schemes, one might obtain different central values for the cross sections and
the impact of this difference should also be addressed\footnote{This difference
should not come as a surprise as, even within the same scheme, there are large
differences when the PDF sets are updated. For instance, as also pointed out in
Refs.~\cite{ggH-FG,ggH-radja}, $\sigma^{\rm NNLO} (gg \to H)$ evaluated with
the MSTW2004 set is different by more than 10\% compared to the current value
obtained with the MSTW2008 set, as a result of a corrected treatment of the
$b,c$ densities among other improvements.}.

In our analysis, we will take into account these two aspects and investigate
the PDF uncertainties given separately by the three MSTW, ABKM and CTEQ
schemes, but we also compare the possibly different central values given by the
three schemes.  Note that despite of the fact that the CTEQ collaboration does
not yet provide PDF sets at NNLO, one can still use the available NLO sets,
evaluating the PDF errors on the NLO cross sections and take these errors as
approximately valid at NNLO, once the cross sections are properly rescaled by
including the NNLO corrections. For the sake of error estimates, this procedure
should provide a good approximation. 

In the case of the  $gg \to H$ cross section at the Tevatron, the  90\% CL  PDF
errors using the three schemes discussed above are shown in Fig.~6 as a
function of $M_H$.  The spread of the cross section due to the PDF errors is
approximately the same in the MSTW and CTEQ schemes, leading to an uncertainty
band of less than $10$\% in both cases. For instance, in the MSTW scheme and in
agreement with Refs.~\cite{ggH-FG,ggH-radja}, we obtain a $\sim \pm 6\%$ error
for $M_H=120$ GeV and $\sim \pm 9\%$ for $M_H=180$ GeV; the errors are only
slightly asymmetric and for $M_H=160$ GeV, one has  $\Delta \sigma^+_{\rm
PDF}/\sigma=+8.1\%$ and $\Delta \sigma^-_{\rm PDF}/\sigma=-8.6\%$.  The errors
are relatively smaller in the ABKM case in the entire Higgs mass range and, for
instance, one  obtains a $\Delta \sigma^\pm_{\rm PDF}/\sigma \approx \pm 5\%$
(7\%) error for $M_H=120~(180)$ GeV.

\begin{figure}[!t]
\begin{center}
\epsfig{file=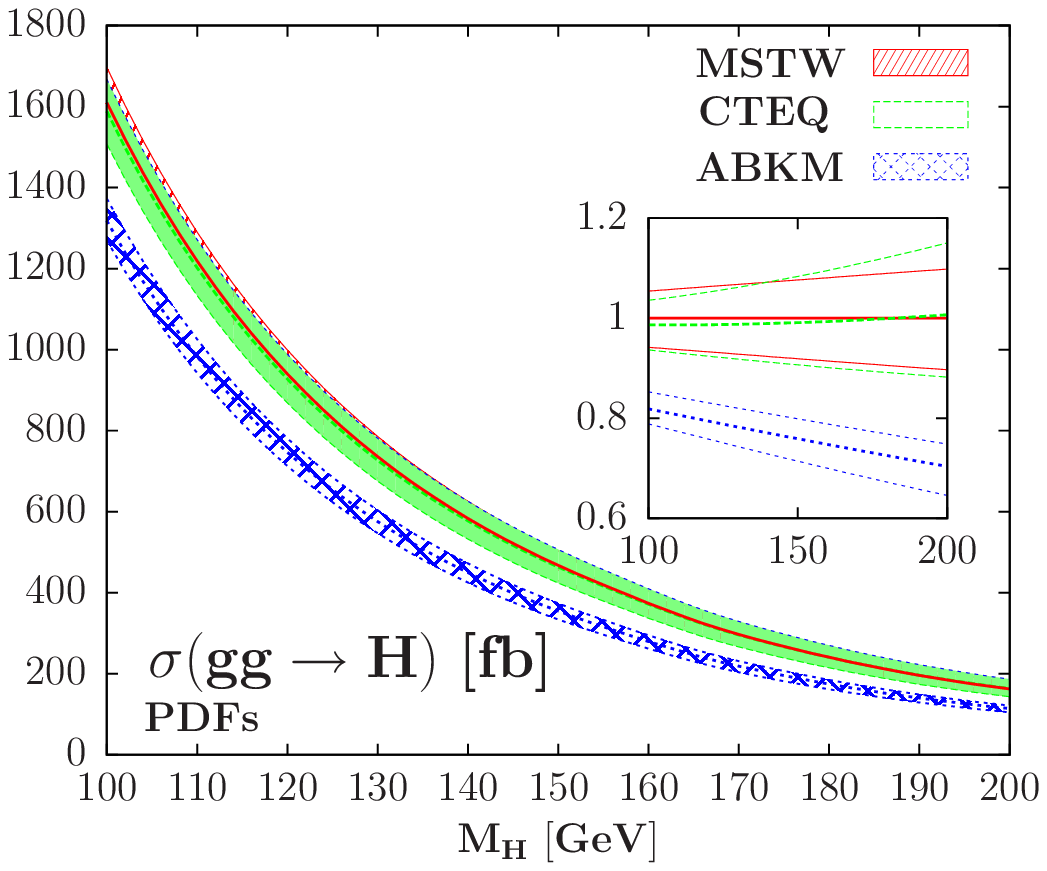,width=11.cm}
\end{center}
\vspace*{-5mm}
\caption[]{The central values and the 90\% CL PDF uncertainty bands in the 
NNLO cross section $\sigma(gg \to H+X)$ at the Tevatron when evaluated within 
the MSTW, CTEQ and ABKM schemes. In the insert, shown in percentage are the  
deviations within a given scheme and the CTEQ and ABKM
central values when the cross sections  are normalized to the MSTW central
value.} 
\vspace*{-2mm}
\end{figure}

A more important issue is the very large discrepancy between the central values
of the cross sections calculated with the MSTW and CTEQ PDFs on the one hand
and the ABKM set of PDFs, on the other hand\footnote{Besides
Refs.~\cite{PDF-as,PDF-Alekhin}, this problem has also been briefly mentioned
in the discussion of Ref.~\cite{Grazinni} which appeared during the final stage
of our work.}. Indeed, the use of the ABKM parametrisation results in a cross
section that is $\sim 25\%$ smaller than the cross section evaluated with the 
MSTW or CTEQ PDFs. Thus, even if the PDF uncertainties evaluated within a given
scheme turn out to be relatively small and apparently well under control, the
spread of the cross sections due to the different parameterisations  can be
much more important. 

If one uses the old way of estimating the PDF uncertainties (i.e. before the
advent  of the PDF error estimates within a given scheme) by comparing  the
results given by different PDF parameterisations, one arrives at an uncertainty 
defined as  
\beq 
\Delta \sigma^+_{\rm PDF}  &=& {\rm max} (  \sigma^0_{\rm MSTW} , \sigma^{0}_{\rm 
CTEQ}, \sigma^0_{\rm ABKM}) -\sigma^0_{\rm MSTW} \nonumber \\
\Delta \sigma^-_{\rm PDF}  &=& \sigma^0_{\rm MSTW} - {\rm min} 
(\sigma^0_{\rm MSTW} , \sigma^{0}_{\rm CTEQ}, \sigma^0_{\rm ABKM})
\label{conservative-PDF} 
\eeq 
where the central value of the $gg\to H$ cross section is taken to be that
given by the MSTW nominal set $S_0$ (we refrain here from adding the
uncertainties obtained  within the same PDF set, which would increase the error
by another  5\% to 7\%). Hence,  for $M_H=160$  GeV for instance, one would
have $\Delta \sigma^+_{\rm PDF} \approx 1\%$ given by the small difference
between the CTEQ and MSTW central values of the cross section and $\Delta
\sigma^-_{\rm PDF} \approx -25\%$ given by the large difference between the
ABKM and MSTW central values. 

However, we would would like to keep considering the MSTW scheme at least for
the fact that it includes the di--jet Tevatron data which are crucial in this 
context. But we would also like understand  the very large  difference in the
$gg\to H$ cross section when evaluated with the MSTW/CTEQ and ABKM sets. This 
difference results not only from the different gluon densities used  (and it is
well known that these densities are less severely constrained by experimental
data than light quark densities), but is also due to the different values of
the strong coupling constant which is fitted altogether with the PDF sets.
Indeed, the value of $\alpha_s$ and its associated error play a crucial role in
the presently discussed production process. For instance, the  $\alpha_s$ value
used in the ABKM set,   $\alpha_s(M_Z^2)=0.1129 \pm 0.0014$ at NLO in the BMSM
scheme \cite{BMSN},  is $\approx 3 \sigma$ smaller than the one in the MSTW set
(see below). Note also that within the dynamical set of  PDFs recently proposed
in Ref~\cite{PDF-JR}, one obtains too an NLO $\alpha_s$  value that is smaller
than the MSTW value but with a slightly  larger uncertainty,
$\alpha_s(M_Z^2)=0.1124 \pm 0.0020$. 

As the $gg \to H$ mechanism is mediated by triangular loops involving the  
heavy top and bottom quarks, the cross section $\sigma(gg \to H)$  is at ${\cal
O}(\alpha_s^2)$ already in the Born approximation and the large NLO and NNLO
QCD contributions are, respectively, of ${\cal O}(\alpha_s^3)$ and ${\cal 
O}(\alpha_s^4)$. Since the corresponding $K$--factors are very large at the
Tevatron, $K_{\rm NLO} \sim 2$ and $K_{\rm NNLO} \sim 3$, a one percent
uncertainty in the input value of $\alpha_s$ will generate a $\approx 3\%$
uncertainty in $\sigma^{\rm NNLO} (gg \to H)$. If, for instance, one uses  
the value of $\alpha_s$ at NLO and its associated experimental uncertainty 
that is fitted in the global analysis of the hard scattering
data performed by the MSTW collaboration \cite{PDF-as}
\beq
\label{alphas}
\alpha_s(M_Z^2)&=& 0.1202~^{+0.0012}_{-0.0015}~{\rm (68\%CL)}~~
                      ^{+0.0032}_{-0.0039}~{\rm (90\%CL)}~~{\rm at~NLO} 
\eeq
leading to $\alpha_s(M_Z^2)=0.1171~^{+0.0014}_{-0.0014}~{\rm (68\%CL)}$ at
NNLO,  by naively plaguing the 90\% CL errors on $\alpha_s$  in the
perturbative series of the partonic cross section but using the best--fit PDF 
set, one arrives at an uncertainty  on the $gg\to H$ cross section that is of
the order of $\Delta \sigma/\sigma  \! \approx\! \pm 8\%$ at the Tevatron, 
over the entire  115 GeV$\lsim M_H \lsim 200$ GeV  range.

Nevertheless, such a naive procedure cannot be applied in practice as, in
general, $\alpha_s$ is fitted together with the PDFs:  the PDF sets are only
defined for the special value of $\alpha_s$ obtained with the best fit and, to
be consistent,  this best value of $\alpha_s$ that we denote $\alpha_s^0$,  
should also be used for the partonic part of the cross section. This adds to
the fact that there is an interplay between the PDFs and the value of
$\alpha_s$ and, for instance, a larger value of $\alpha_s$ would lead to a
smaller gluon density at low $x$ \cite{PDF-as}.

Fortunately enough, the MSTW collaboration released very recently a new
set--up which allows for a simultaneous evaluation of the errors due to the
PDFs and those due to the experimental uncertainties  on $\alpha_s$ of
eq.~(\ref{alphas}), taking into account the possible correlations 
\cite{PDF-as}.  The procedure to obtain the different PDFs and their associated
errors is similar to the one discussed before, but provided is a collection of
five PDF+error sets for different  $\alpha_{s}$ values: the best fit value
$\alpha_s^0$ and its 68\% CL and 90\% CL maximal and minimal values. 
Using the following equations to calculate the PDF error for a fixed value
of $\alpha_{s}$,
\beq
\left(\Delta\sigma_{{\rm PDF}}^{\alpha_{s}}\right)_{+} =\sqrt{\sum_{i}
\left\{\max\left[
 \sigma (\alpha_{s}, S_{i}^{+})-\sigma (\alpha_{s}^0,S_{0}),
 \sigma (\alpha_{s}, S_{i}^{-})-\sigma (\alpha_{s}^0,S_{0}),0 \right] 
 \right\}^2}\, ,
 \label{pdferrorplus} \nonumber \\
\left(\Delta\sigma_{{\rm PDF}}^{\alpha_{s}}\right)_{-} =\sqrt{\sum_{i}
\left\{\max\left[
 \sigma (\alpha_{s}^0,S_{0})-\sigma (\alpha_{s},S_{i}^{+}),
 \sigma (\alpha_{s}^0,S_{0})-\sigma (\alpha_{s},S_{i}^{-}),0 \right]
 \right\}^2}\, ,
 \label{pdferrorminus}
    \eeq
one then  compares these five different values and finally arrives,  with
$\alpha_{s}^0$ as the best--fit value of $\alpha_{s}$ given by the central 
values of eq.~(\ref{alphas}) and $S_0$ the nominal PDF set with this 
$\alpha_s$ value, at the 90\% CL PDF+$\Delta^{\rm exp} \alpha_s$  errors given 
by \cite{PDF-as}
    \beq
\Delta\sigma^+_{{\rm PDF}+\alpha^{\rm exp}_{s}} = \max_{\alpha_{s}}\left( 
\left\{\sigma (\alpha_{s}^0,S_{0})+
\left(\Delta\sigma_{{\rm PDF}}^{\alpha_{s}}\right)_{+}  \right\}\right)
-\sigma (\alpha_{s}^{0},S_{0}) \, ,
 \label{pdfaserrorplus} \nonumber \\
\Delta\sigma^-_{{\rm PDF}+\alpha^{\rm exp}_{s}} =\sigma (\alpha_{s}^{0},S_{0})-
\min_{\alpha_{s}}\left( \left\{\sigma (\alpha_{s}^0,S_{0})-
\left(\Delta\sigma_{{\rm PDF}}^{\alpha_{s}}\right)_{-}  \right\}\right)\, .
 \label{pdfaserrorminus}
    \eeq

Using this procedure, we have evaluated the PDF+$\Delta^{\rm exp} \alpha_s$
uncertainty on the NNLO $gg \to H$ total cross section at the Tevatron and  the
result is displayed in the left--hand side of Fig.~7 as a function of $M_H$.
The PDF+$\Delta^{\rm exp} \alpha_s$ error ranges from $\approx \pm 11\%$ for
$M_H=120$ GeV to $ \approx \pm 14\%$ for $M_H=180$ GeV with, again, a slight
asymmetry between the upper and lower values; for a Higgs mass  $M_H=160$ GeV,
one has $\Delta\sigma^\pm_{ {\rm PDF}+\alpha_{s}} /\sigma = ^{+12.8\%}_{-
12.0\%}$.   That is, the experimental uncertainty on $\alpha_s$  adds a
$\approx 5\%$ error to the PDF error alone over the entire $M_H$ range relevant
at the Tevatron. This is a factor of $\approx 1.5$  less than the naive guess 
made previously,  as a   result of the correlation between the PDFs and the
$\alpha_s$ value. 

Nevertheless, this larger PDF+$\Delta^{\rm exp} \alpha_s$ uncertainty  compared
to the PDF uncertainty alone does not yet reconcile the evaluation of MSTW and
ABKM (in this last scheme the $\Delta^{\rm exp}\alpha_s$ uncertainty has not
been  included  since no PDF set with an error on $\alpha_s$ is provided) 
of the $gg\to H$ cross section at the Tevatron, the difference between the
lowest MSTW value and the highest ABKM value being still at the level of
$\approx  10\%$. 

\begin{figure}[!h]
\begin{center}
\vspace*{3mm}
\epsfig{file=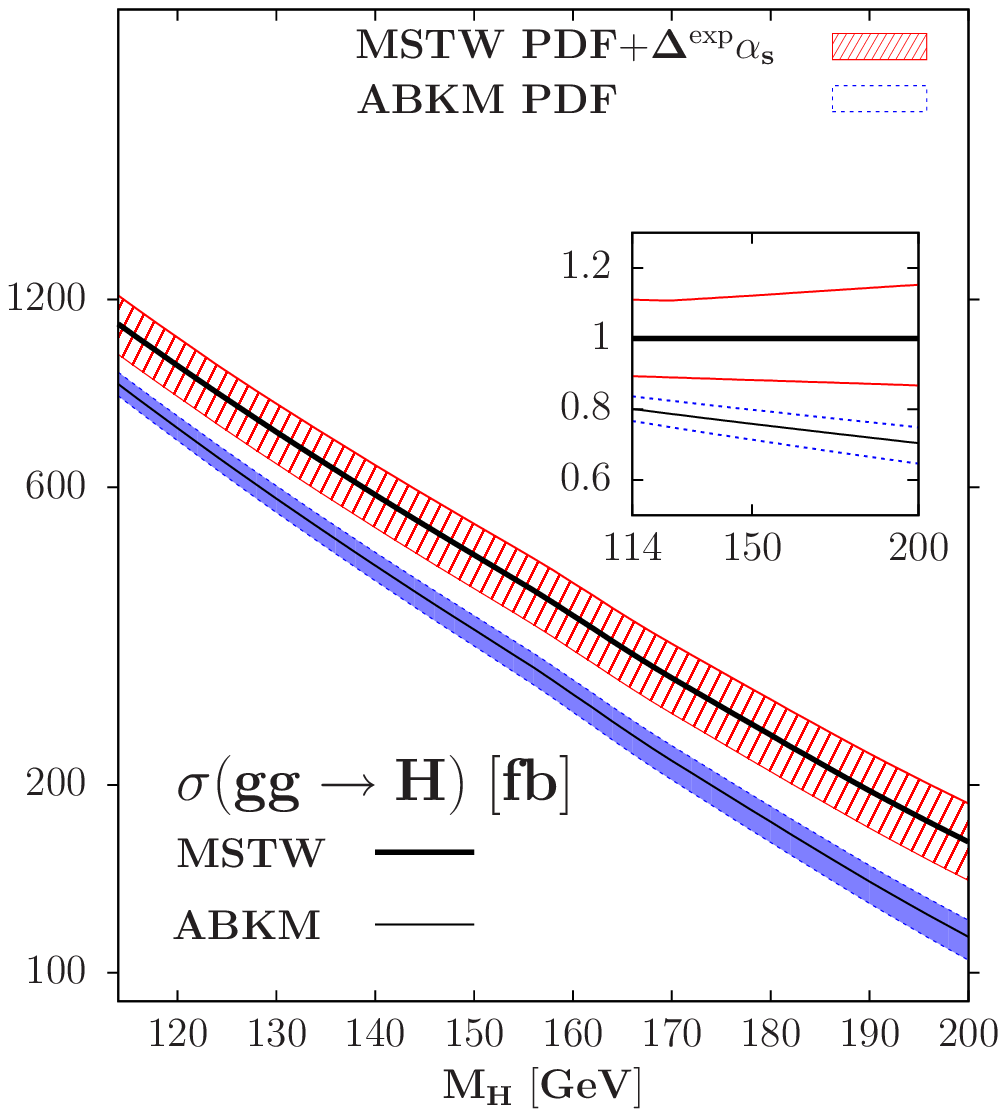,width=7.5cm}
\epsfig{file=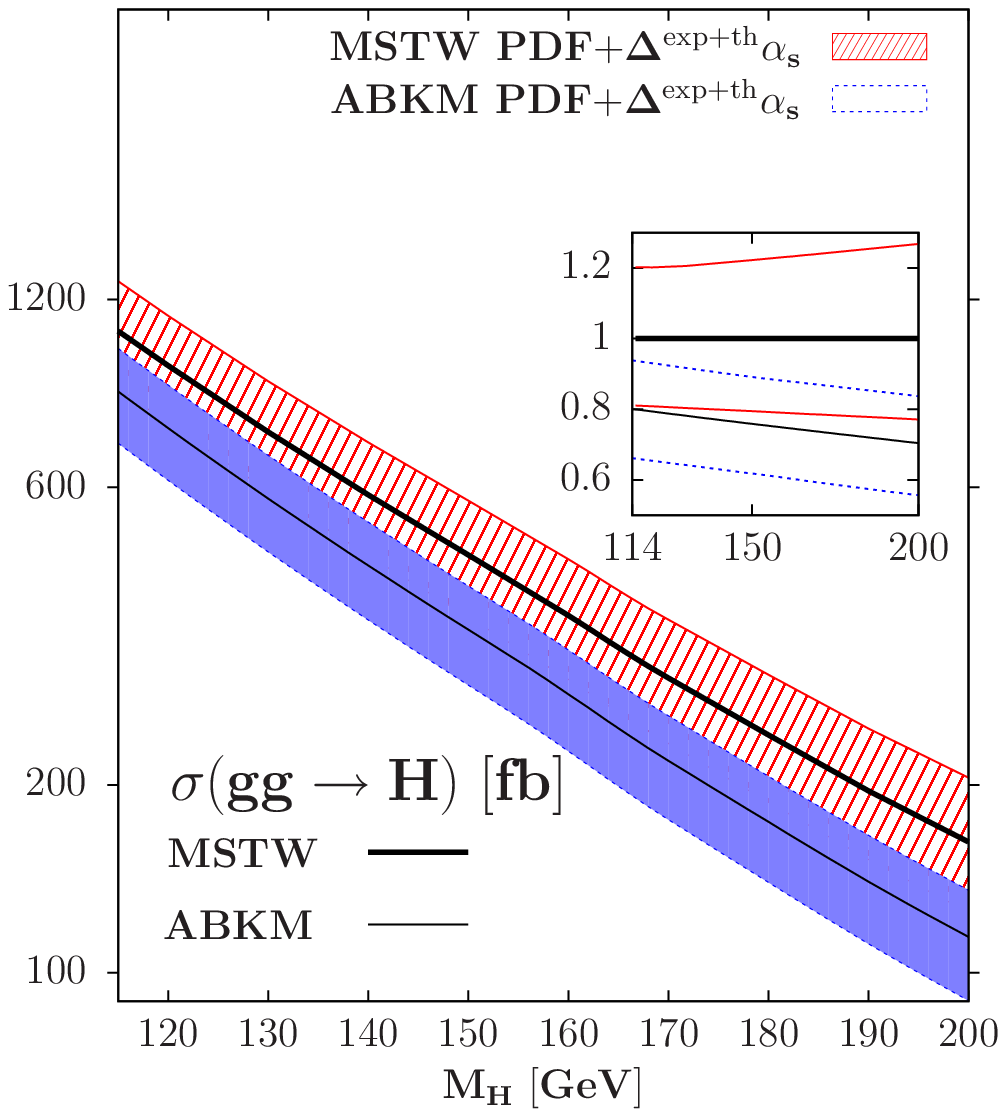,width=7.5cm}
\end{center}
\vspace*{-3mm}
\caption[]{
Left: the  PDF+$\Delta^{\rm exp}\alpha_s$ uncertainties in the MSTW scheme  
and the PDF uncertainties in the ABKM schemes on the $gg \to H$ cross section
at the Tevatron as a function of $M_H$.  Right: the  PDF+$\Delta^{\rm exp}\alpha_s
+\Delta^{\rm th}\alpha_s$ uncertainties in the MSTW scheme using the new set--up
and the PDF+$\Delta^{\rm exp}\alpha_s$+$\Delta^{\rm th}\alpha_s$ error in the ABKM 
scheme using our naive procedure.  In the inserts, shown are the same but with 
the cross sections normalized to the MSTW central cross section.}
\end{figure}

So far, only the impact of the experimental errors on $\alpha_s$ has been
discussed, while it is well known that the strong coupling constant is also
plagued by theoretical uncertainties due to scale variation, ambiguities in
heavy quark flavor scheme definition, etc..  In Ref.~\cite{PDF-MSTW} this
theoretical error has been estimated to be at least $\Delta^{\rm th} \alpha_s=
\pm 0.003$ at NLO ($\pm 0.002$ at NNLO) while the estimate of
Ref.~\cite{PDF-Alekhin-old} leads to a slightly larger uncertainty,
$\Delta^{\rm th} \alpha_s= \pm 0.0033$. Unfortunately, this theoretical error
is not taken into account in the MSTW PDF+$\Delta \alpha_s$ error set--up
discussed above, nor is addressed by any  of the other  PDF schemes. 

Adopting the smallest of the $1\sigma$  $\alpha_s$ errors at NLO quoted above,
i.e. 
\beq
\Delta^{\rm th} \alpha_s= 0.003 \, , 
\label{alphas_th}
\eeq
we have evaluated the uncertainty due this theoretical error on $\sigma ^{\rm
NNLO} (gg \to H+X)$ at the Tevatron, following our naive and admittedly  not
entirely consistent first estimate of the impact of the experimental error  of
$\alpha_s$ on the same cross section, i.e. using the values $\alpha_s^0 \pm
0.002$ in the partonic cross sections but the best--fit value $\alpha_s^0$ in
the best--fit PDF set.  We obtain an error of  $\approx 8\%$ on $\sigma^{\rm
NNLO}  (gg\! \to\! H)$ for the $M_H$ values relevant at the Tevatron.

There is nevertheless a more consistent way to address this issue of the 
theoretical uncertainty on $\alpha_s$, thanks to a fixed--$\alpha_s$ NNLO PDF
grid also provided by the MSTW collaboration, which is a set of central PDFs
but at  fixed values of $\alpha_s$ different from the best--fit value. Values
of  $\alpha_s$ in a range comprised between $0.107$ and $0.127$ in steps of
0.001 are selected, and thus include the values $\alpha_s^0\pm 0.002$ that are
interesting for our purpose. Using this PDF grid with the theoretical error on 
$\alpha_s$ of eq.~(\ref{alphas_th}) implemented, the upper and lower values 
of the cross sections will be  given by
\beq
\Delta \sigma^{+}_{\rm PDF+\alpha_s^{th}} = \sigma (\alpha_s^{0}+ \Delta^{\rm th} 
\alpha_s, S_0(\alpha_s^{0}+\Delta^{\rm th} \alpha_s)) - \sigma (\alpha_s^0, 
S_0(\alpha_s^{0})) \nonumber \\
\Delta \sigma^-_{\rm PDF+\alpha_s^{ th}} = \sigma (\alpha_s^{0}, S_0(\alpha_s^0 )) 
-\sigma (\alpha_s^{0}- \Delta^{\rm th} \alpha_s, S_0(\alpha_s^{0}- \Delta^{\rm 
th} \alpha_s))
\label{as-naive}
\eeq
with again $S_0(\alpha_s)$ being the MSTW best--fit PDF set at the fixed
$\alpha_s$ value which is either $\alpha_{s}^0$ or $\alpha_{s}^0
\pm\Delta^{\rm  th}\alpha_s$. With this fixed--$\alpha_s$ PDF grid, we obtain
an error of  $\approx  + 10\%$ and  $\approx -9\%$ on the total $gg \to H$
cross section at NNLO when  one restricts to  the range of Higgs masses
relevant at the Tevatron, with a $\approx 1\%$ increase from $M_H=115$ GeV to 
$M_H=200$ GeV. This error is again very close  to the naive  estimate performed
previously by considering only the impact of $\Delta^{\rm th}\alpha_s$  on the
partonic cross section. Note that despite of the fact that the uncertainty on
$\alpha_s$ is a theoretical one and is not at the 90\% CL, we will take  the
PDF+$\Delta^{\rm th}\alpha_s$ error that one obtains using the equations above 
to be at the 90\% CL.

In the MSTW scheme, to obtain the total PDF+$\alpha_s$ uncertainty, one then 
adds  in quadrature the PDF+$\Delta^{\rm exp}\alpha_s$ and PDF+$\Delta^{\rm th}
\alpha_s$ uncertainties, 
\beq
\Delta \sigma^\pm_{\rm PDF+\alpha_s^{exp}+\alpha_s^{th}}&=&
\left( (\Delta \sigma^\pm_{\rm PDF+\alpha_s^{exp}})^2+
       (\Delta \sigma^\pm_{\rm PDF+\alpha_s^{th} })^2 \right)^{1/2} . 
\label{as-combined}
\eeq

The result for the total PDF+$\alpha_s$ 90\% CL uncertainty on $\sigma^{\rm
NNLO}( gg  \to H)$ in the MSTW scheme using the procedure outlined above is
shown in the  right--hand side of Fig.~7 as a function of $M_H$. It is  
compared to the result when the PDF error in the ABKM scheme is combined with
the $\Delta^{\rm  exp}\alpha_s$ and $\Delta^{\rm  th}\alpha_s$ uncertainties
using the naive procedure discussed previously as, in this case, no PDF with an
$\alpha_s$ value different from that obtained with the best--fit is provided. 
One can see that the results given by the two parameterizations appear now to
be consistent with each other as the two uncertainty bands overlap. 

The net result of this exercise is that the total error on the $gg\to H$ cross
section due to the PDF and  the theoretical plus experimental uncertainties on
$\alpha_s$, is now rather significant and, in the case of the MSTW scheme to
which we stick, it amounts to approximately $\pm 15$ to 20\% in the Higgs mass
range relevant at the Tevatron. The uncertainty is, for instance,  $- 15\%$ and
$+16.5\%$ for $M_H=160$ GeV and is substantially smaller (for the minimal value
of the cross section) than the error that would have been obtained using the
old--fashioned  estimate of the PDF errors by comparing different PDF sets, in
which case one would have had an uncertainty of  $-26\%$ and $+1\%$ compared to
the MSTW central value.

  The final error of $\approx \pm 15$--20\%  is to be compared to the $\pm 
6$--10\% error obtained from the PDF uncertainty alone ($\approx \pm 8\%$ for
$M_H=160$ GeV), an amount which has been taken to be the total PDF uncertainty
in the CDF/D0 analysis of the Higgs signal. Thus, similarly to the scale
variation,  the PDF uncertainties, when the errors on $\alpha_s$ are  taken
into account, have been underestimated by at least a  factor of two by the
experiments.

\section{Theoretical uncertainties in Higgs--strahlung}

We now turn to the discussion of the theoretical uncertainties in the Higgs
strahlung mechanism $q\bar q \to VH$, following the same line of arguments as
in the previous section. Since in this case, the NNLO QCD corrections and the
one--loop electroweak corrections have been obtained exactly and no effective
approach was used, only the scale variation and the PDF+$\alpha_s$
uncertainties have to be discussed. In addition, since the NNLO gluon--gluon
fusion contribution to the cross section in the $p \bar p \to ZH$ case, which
is absent in $p \bar p \to WH$, is very small at the Tevatron and
because the scales and phase space are only slightly different for the $p \bar p
\to WH$ and $ZH$ processes, as the difference $(M_Z^2-M_W^2)/ \hat s$ is tiny,
the kinematics and the $K$--factors for these two processes are very similar.
We thus restrict our analysis to the $WH$ channel but the same results hold for
the $ZH$ channel. 

To evaluate the uncertainties due to the variation of the renormalisation and 
factorisation scales in the Higgs--strahlung processes, the choice of the 
variation domain is in a sense simpler than for the $gg\to H$ mechanism.
Indeed, as the process at leading order is mediated solely by massive gauge
boson exchange and, thus, does not involve strong interactions at the partonic
level, only the factorisation scale $\mu_F$ appears when the partonic cross
section is folded with the $q$ and $\bar q$ luminosities and there is no
dependence on the renormalisation scale $\mu_R$ at this order. It is only at
NLO, when gluons are exchanged between or radiated from the $q,\bar q$ initial
states, that both scales $\mu_R$ and $\mu_F$ appear explicitly.

 Using our proposed criterion for the estimate of the perturbative higher order
effects, we thus choose again to consider the variation domain of the scales
from their central values, $\mu_0/\kappa \le \mu_R, \mu_F \le \kappa \mu_0$
with $\mu_0=M_{HW}$, of the NLO cross section instead of that of the LO cross
section to determine the value of the factor $\kappa$ to be used at NNLO.  We
display in the left-hand side of Fig.~8 the variation of the NLO cross section
$\sigma^{\rm NLO} (p \bar p \to WH)$ at the Tevatron as a function of $M_H$ for
three values of the constant $\kappa$  which defines the range spanned by the
scales, $M_{HW}/\kappa \le \mu_R, \mu_F \le \kappa M_{HW}$. One sees that, in
this case, a value $\kappa=2$ is sufficient (if the scales $\mu_R$ and $\mu_F$
are varied independently in the chosen domain) in order that the uncertainty
band at NLO reaches the central value of the cross section at NNLO. In fact,
the NLO uncertainty band would have been only marginally  affected if one had
chosen the values $\kappa=3$, $4$ or even 5. This demonstrates than the cross
sections for the Higgs--strahlung processes, in contrast to $gg \to H$, are
very stable against scale variation, a result that is presumably due to the
smaller $q\bar q$ color charges compared to gluons, $\approx  C_F/C_A$,  that
lead to more moderate QCD corrections.

In the right--hand side of Fig.~8, the NNLO $p\bar p \to WH$ total cross section
is displayed as a function of $M_H$ for a  scale variation $\frac12 M_{HW} \le
\mu_R, \mu_F  \le 2 M_{HW}$. Contrary to the $gg \to H$ mechanism, the scale
variation within the chosen range is rather mild and only  a $\sim 0.7\%$ (at
low $M_H$) to $1.2\%$ (at high $M_H$) uncertainty is observed for the relevant 
Higgs mass range at the Tevatron. This had to be expected as the $K$--factors in
the Higgs--strahlung processes, $K_{\rm NLO} \approx 1.4$ and  $K_{\rm NNLO}
\approx 1.5$,  are substantially smaller than those affecting the $gg$ fusion
mechanism and one expects perturbation theory to have a better behavior in the
former case. This provides more confidence that the Higgs--strahlung cross
section is stable against scale variation and, thus, that higher order effects
should be small.

\begin{figure}[!h]
\begin{center}
\epsfig{file=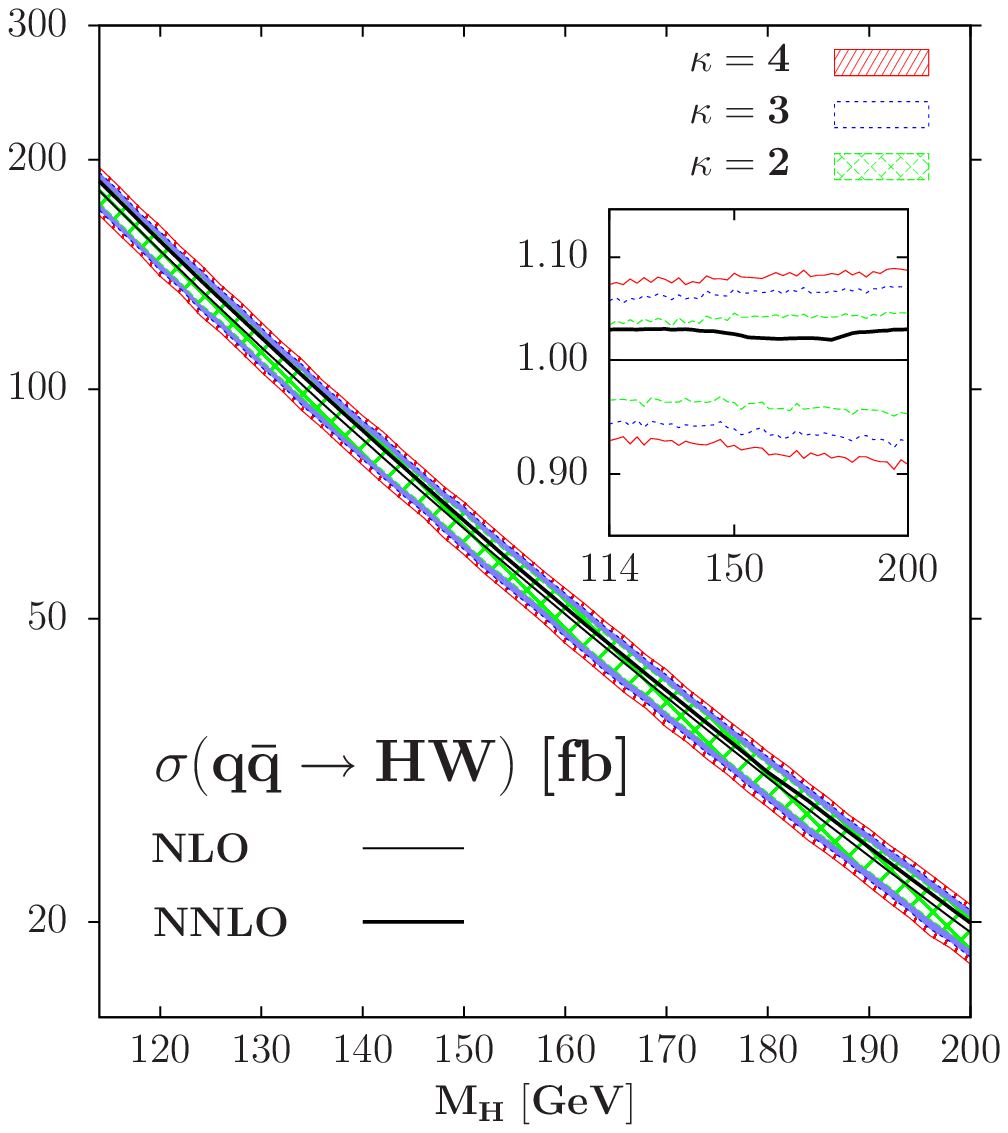,width=7.cm}
\epsfig{file=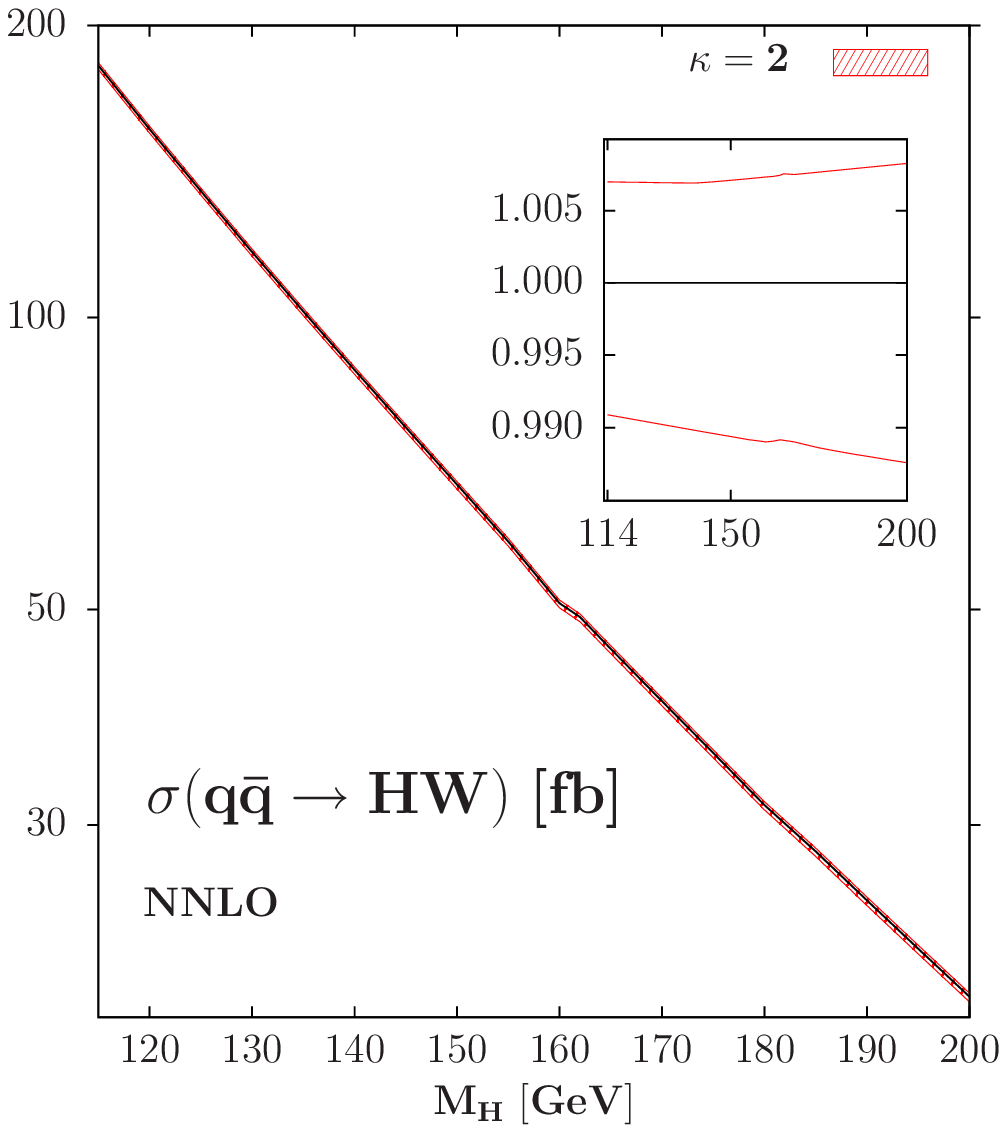,width=7.cm} 
\end{center}
\vspace*{-3mm}
\caption[]{Left: the scale dependence of $\sigma(p \bar p \to WH)$  at 
NLO for variations $M_{HV}/{\kappa} \le \mu_R, \mu_F \le \kappa M_{HV}$ with 
$\kappa=2,3$ and 4, compared to the NNLO value; in the insert, shown are the 
variations in percentage  and where the NNLO cross section is normalized to the NLO
one. Right: the scale  dependence of $\sigma( p\bar p \to WH)$ at NNLO for a
variation in the domains  $M_{HV}/2 \le \mu_R= \mu_F \le 2 M_{HV}$; the relative
deviations from the  central value are shown in the insert.}
\vspace*{-2mm}
\end{figure}

For the estimate of the uncertainties due to the PDFs in associated Higgs 
production with a $W$ boson, $p \bar p \to WH$ (again, the output is similar
for $p\bar p \to ZH$  except from the overall normalisation, despite of the 
different initial state (anti)quarks), the same exercise made in section 3.3
for the $gg$ fusion mechanism has been repeated. The results are shown in
Figs.~9 and 10 for Tevatron energies as a function of $M_H$. Figure 9 displays
the spread of the $p\bar p \to WH$ cross section due to the PDF uncertainties
alone in the MSTW, CTEQ and ABKM schemes and, again in this case, the
uncertainty bands are similar in the CTEQ and MSTW schemes and lead to an error
of about 4\%;  the band is, however, slightly larger in the ABKM scheme. Here
also appears a discrepancy between the MSTW/CTEQ  and the ABKM central values,
the cross section with the PDFs from ABKM being this time about 10\% larger
than that obtained with the other sets. However, in contrast to  the $gg\to
H$ case, the MSTW/CTEQ and ABKM uncertainty bands almost touch each other.

\begin{figure}[!h]
\begin{center}
\epsfig{file=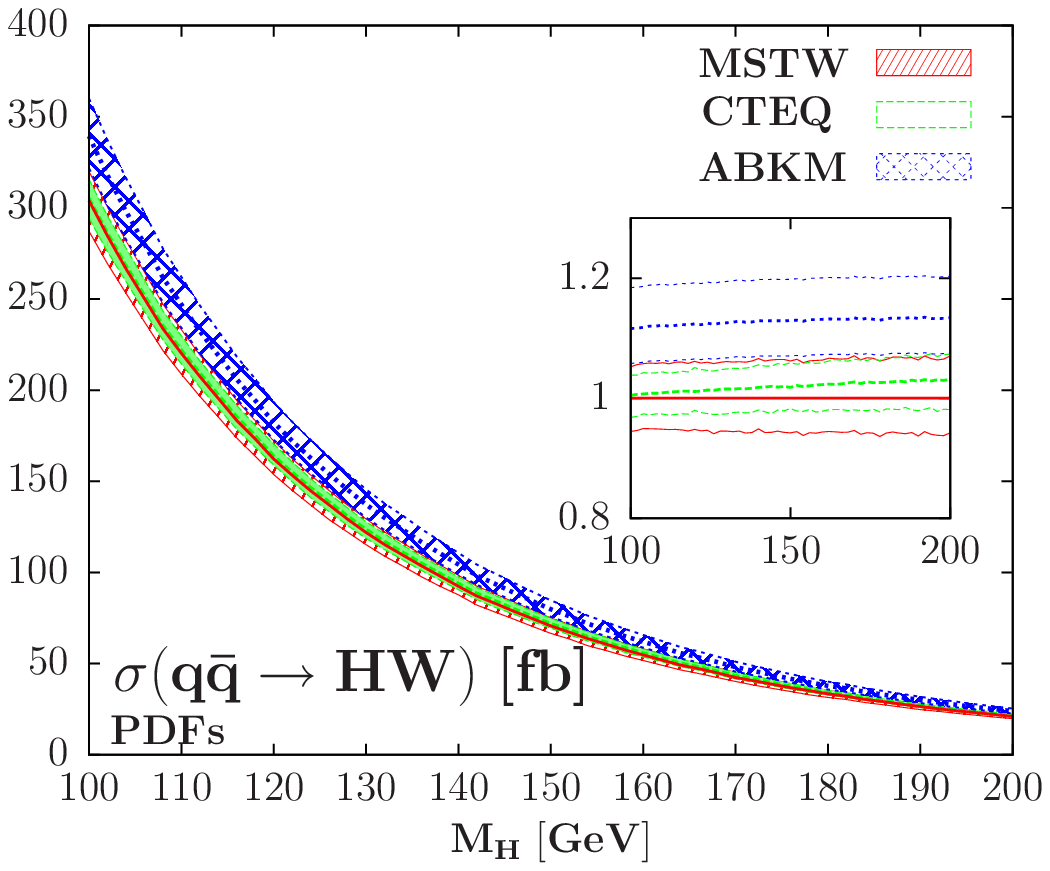,width=10.cm}
\end{center}
\vspace*{-2mm}
\caption[]{The central values and the PDF uncertainties in the cross 
section $\sigma( p \bar p \to WH)$ at the Tevatron when evaluated within the 
MSTW, CTEQ and ABKM schemes. In the insert, the relative deviations from the 
central MSTW value are shown.} 
\vspace*{-2mm}
\end{figure}

\begin{figure}[!h]
\vspace*{5mm}
\begin{center}
\epsfig{file=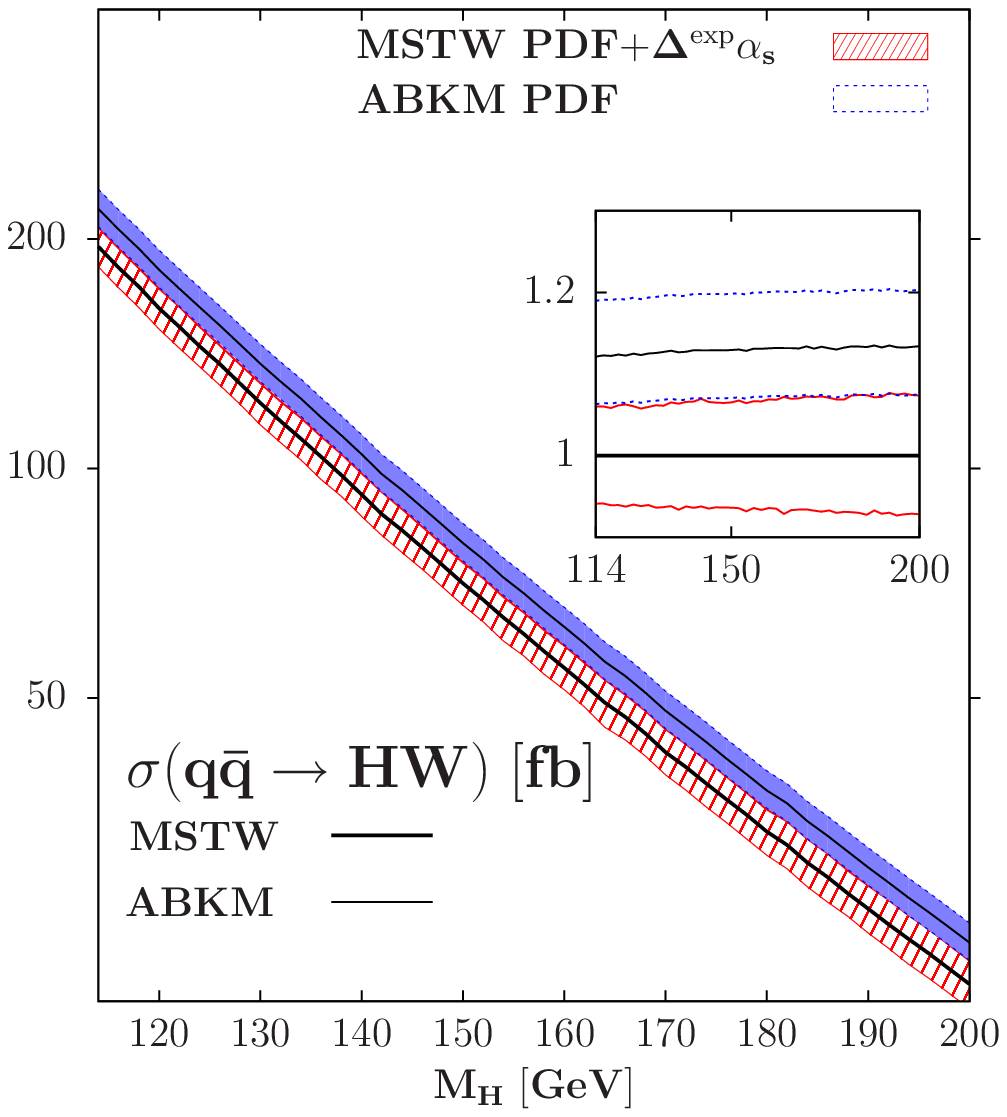,width=7.cm}
\epsfig{file=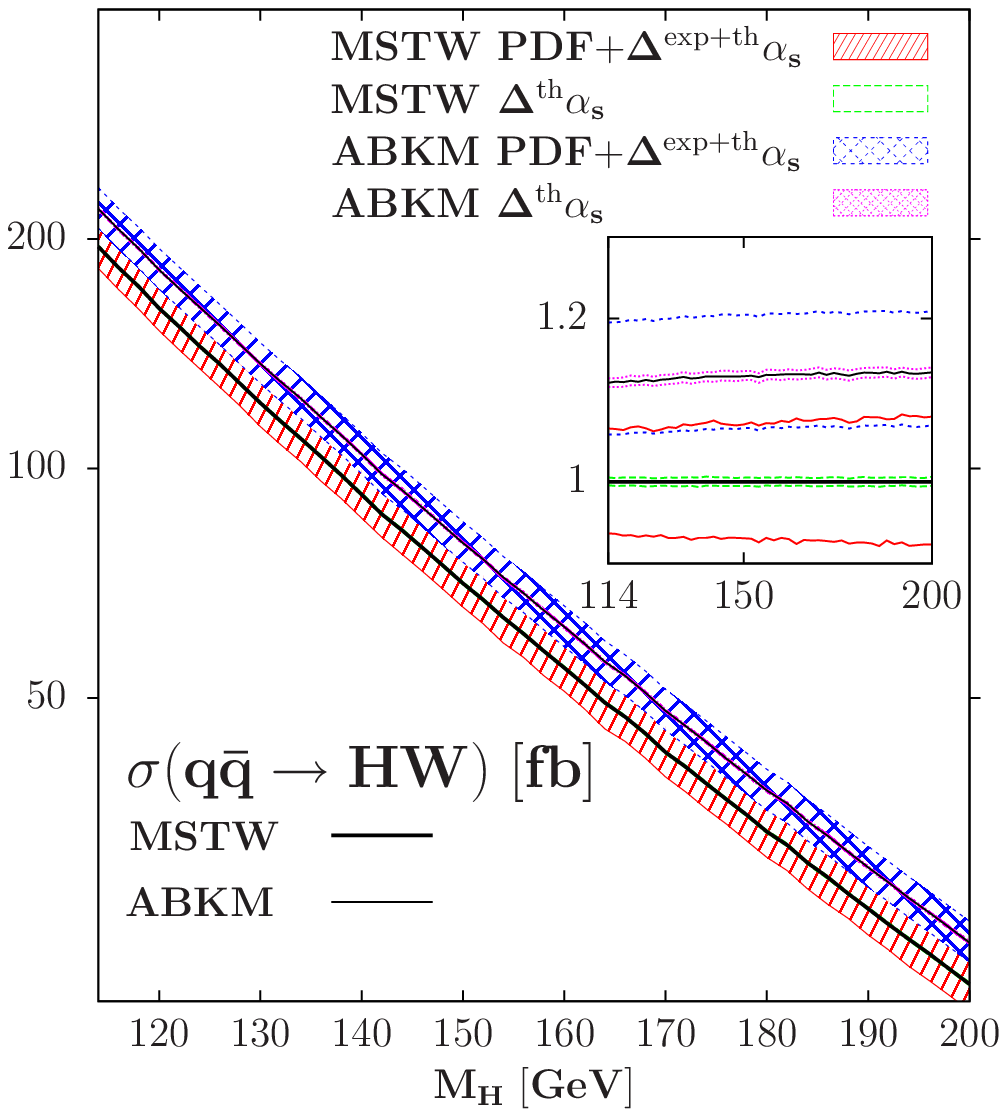,width=7.cm}
\end{center}
\vspace*{-5mm}
\caption[]{Left: the  PDF uncertainties in the MSTW and ABKM schemes when the
additional experimental errors on $\alpha_s$ is included in MSTW  as discussed
in  the text; in the insert, the relative deviations from the  central MSTW
value are shown. Right: the same as in a) but when the  theoretical error on
$\alpha_s$ is added in both the MSTW and ABKM cases. }  
\vspace*{-2mm}
\end{figure}

In the left--hand side of Fig.~10, we show the  bands resulting from the
PDF+$\Delta^{\rm exp} \alpha_s$ uncertainty in the MSTW mixed scheme, while the
right--hand side of the figure shows the  uncertainty bands when the 
additional  theoretical error  $ \Delta^{\rm th} \alpha_s$ is included in both
the MSTW scheme using eq.~(\ref{as-combined}) and  ABKM scheme using  the 
naive estimate of eq.~(\ref{as-naive}). As expected, the errors due to the
imprecise value of $\alpha_s$ are much smaller than in the $gg \to H$
mechanism, as in Higgs--strahlung, the process does not involve $\alpha_s$ in
the Born approximation and the $K$--factors are reasonably small, $K_{\rm NNLO}
\lsim 1.5$. Hence, $\Delta^{\rm exp}\alpha_s$ generates an additional error
that is about $\approx  2\%$  when included in the PDF fits, while the error
due to  $\Delta^{\rm th}\alpha_s$ is about one to two percent. 

Nevertheless, the total PDF+$\Delta^{\rm exp}\alpha_s$+$\Delta^{\rm
th}\alpha_s$ uncertainty is at the level of $\approx 7$--8\% in the MSTW
scheme, i.e. slightly larger than the errors due to the PDFs alone, and
arranges again  so that the MSTW and ABKM uncertainty bands have a significant
overlap.  

\section{The total uncertainties at the Tevatron} 

The analysis of the Higgs production cross section in the $gg \to H$  process
at the Tevatron, as well as the various associated theoretical uncertainties,
is summarized in Table 2. For a set of Higgs mass values that is relevant at
the Tevatron (we choose a step of 5 GeV as done by the CDF and D0 experiments
\cite{Tevatron} except in the critical range 160--170 GeV where a 2 GeV step 
is adopted), the second column of the table gives the central values of the
total cross section at NNLO (in fb) for the renormalisation and factorisation
scale choice $\mu_R=\mu_F=M_H$,  when the partonic cross sections are folded
with the MSTW parton densities. The following columns give the errors on the
central value of the cross section originating from the various sources
discussed in section 3, namely, the uncertainties due to the scale variation in
the adopted range $\frac13 M_H \le  \mu_R, \mu_F \le 3 M_H$, the 90\% CL
errors  due to the MSTW PDF, PDF+$\Delta^{\rm  exp}\alpha_s$ and
PDF+$\Delta^{\rm  exp}\alpha_s$+$\Delta^{\rm th}\alpha_s$ uncertainties as well
as  the estimated uncertainties from the use of the effective approach in the
calculation of the NNLO QCD (the $b$--quark loop contribution and its
interference with the top--quark loop) and electroweak (difference between the
complete and partial factorisation approaches) radiative corrections. 

The largest of these errors, $\sim 20\%$, is due to the scale variation,
followed by the PDF+$\Delta^{\rm exp} \alpha_s +\Delta^{\rm th}\alpha_s$
uncertainties which are at the level of $\approx 15\%$; the errors due to  the
effective theory approach (including that due to the definition of the
$b$--quark mass) are much smaller, being of the order of a few percent for both
the QCD and electroweak parts.

The next important issue is how to combine these various uncertainties. In
accord with Ref.~\cite{ggH-ADGSW}, we do not find any obvious justification to
add these errors in quadrature as done, for instance, by the CDF and D0
collaborations\footnote{In earlier analyses, the CDF collaboration
\cite{HWW-CDF,HWW-Greg} adds in quadrature the 10.9\% scale uncertainty
obtained at NNLL with a scale variation in the range $\frac12 M_H \le  \mu_R,
\mu_F \le 2 M_H$ with a 5.1\% uncertainty due the errors on the MSTW PDFs (not
including the errors from $\alpha_s$), resulting in a 12\% total uncertainty.
The D0 collaboration \cite{HWW-D0,HWW-Greg}  assigns an even smaller total
error, 10\%, to the production cross section.} \cite{HWW-CDF,HWW-D0,HWW-Greg}.
Indeed, while the PDF+$\alpha_s$ uncertainty might have some statistical
ground, the scale uncertainty as well as the uncertainties due to the use of
the effective approach are purely theoretical errors. On the other hand, one
cannot simply add these errors linearly as is generally done for  theoretical
errors, the reason being  a possibly strong interplay between the scale chosen
for the process, the value of $\alpha_s$ (which evolves with the scales) and
thus the PDFs (since the gluon density, for instance, is sensitive to the exact
value of $\alpha_s$ as  mentioned previously). Here, we propose a simple
procedure to  combine at least the two largest uncertainties, those due to the
scale variation and to the PDF+$\alpha_s$ uncertainties, that is in our opinion
more adequate and avoids the drawbacks of the two other possibilities
mentioned above.

The procedure that we propose is as follows. One first derives the maximal and
minimal values  of the production cross sections when the renormalisation and
factorisation scales are varied in the adopted  domain, that is, $\sigma_0 \pm
\Delta \sigma^\pm_\mu$ with  $\sigma_0$ being the cross section evaluated for
the central scales $\mu_R=\mu_F=\mu_0$ and the deviations  $\Delta
\sigma^\pm_\mu$  given in eq.~(3.2). One then evaluates on these maximal and
minimal cross sections from scale variation, the PDF+$\Delta^{\rm exp}
\alpha_s$ as well as the PDF+$\Delta^{\rm th} \alpha_s$ uncertainties (combined
in quadrature) using the new MSTW set-up, i.e  as in eq.~(\ref{as-combined})
but with $\sigma_0$ replaced by $\sigma_0 \pm \Delta \sigma_\mu^\pm$.  

\TABLE[!h]{\small%
\let\lbr\{\def\{{\char'173}%
\let\rbr\}\def\}{\char'175}%
\renewcommand{\arraystretch}{1.66}
\vspace*{10mm}
\begin{tabular}{|c||c||c|ccc|cc||cc|}\hline
$~~M_H~~$ & $\sigma^{\rm NNLO}_{\rm gg \to H}$ [fb]&~~scale~~&PDF&PDF+$
\alpha_s^{\rm exp}$&$\alpha_s^{\rm th}$ & EW & b--loop & total & \% total \\ \hline
$115$ & 1068 & $^{+244}_{-226}$ & $^{+65}_{-69}$ & $^{+118}_{-113}$ & $^{+97}_{-90}$ &
$^{+32}_{-32}$ & $^{+28}_{-28}$ & $^{+507}_{-416}$ & $^{+47\%}_{-39\%}$ \\  \hline
$120$ & 940 & $^{+212}_{-199}$ & $^{+59}_{-63}$ & $^{+103}_{-101}$ & $^{+87}_{-79}$ &
$^{+29}_{-29}$ & $^{+25}_{-25}$ & $^{+446}_{-368}$ & $^{+47\%}_{-39\%}$ \\  \hline
$125$ & 830 & $^{+185}_{-176}$ & $^{+54}_{-58}$ & $^{+90}_{-90}$ & $^{+78}_{-71}$ &
$^{+27}_{-27}$ & $^{+21}_{-21}$ & $^{+394}_{-327}$ & $^{+47\%}_{-39\%}$ \\  \hline 
$130$ & 736 & $^{+163}_{-156}$ & $^{+50}_{-53}$ & $^{+82}_{-81}$ & $^{+70}_{-63}$ &
$^{+24}_{-24}$ & $^{+18}_{-18}$ & $^{+349}_{-291}$ & $^{+47\%}_{-40\%}$ \\  \hline
$135$ & 654 & $^{+144}_{-139}$ & $^{+46}_{-49}$ & $^{+74}_{-73}$ & $^{+63}_{-56}$ &
$^{+22}_{-22}$ & $^{+16}_{-16}$ & $^{+312}_{-260}$ & $^{+48\%}_{-40\%}$ \\  \hline 
$140$ & 584 & $^{+128}_{-124}$ & $^{+42}_{-45}$ & $^{+68}_{-66}$ & $^{+57}_{-51}$ &
$^{+21}_{-21}$ & $^{+14}_{-14}$ & $^{+279}_{-234}$ & $^{+48\%}_{-40\%}$ \\  \hline
$145$ & 522 & $^{+113}_{-111}$ & $^{+39}_{-41}$ & $^{+62}_{-60}$ & $^{+52}_{-46}$ &
$^{+19}_{-19}$ & $^{+12}_{-12}$ & $^{+250}_{-209}$ & $^{+48\%}_{-40\%}$ \\  \hline 
$150$ & 468 & $^{+101}_{-99}$ & $^{+36}_{-38}$ & $^{+57}_{-55}$ & $^{+47}_{-41}$ &
$^{+17}_{-17}$ & $^{+11}_{-11}$ & $^{+225}_{-188}$ & $^{+48\%}_{-40\%}$ \\  \hline
$155$ & 419 & $^{+90}_{-89}$ & $^{+33}_{-35}$ & $^{+52}_{-50}$ & $^{+43}_{-37}$ &
$^{+15}_{-15}$ & $^{+9}_{-9}$ & $^{+202}_{-169}$ & $^{+48\%}_{-40\%}$ \\  \hline 
$160$ & 374 & $^{+79}_{-80}$ & $^{+30}_{-32}$ & $^{+48}_{-45}$ & $^{+39}_{-34}$ &
$^{+11}_{-11}$ & $^{+8}_{-8}$ & $^{+178}_{-149}$ & $^{+48\%}_{-40\%}$ \\  \hline
$162$ & 357 & $^{+76}_{-76}$ & $^{+29}_{-31}$ & $^{+46}_{-43}$ & $^{+37}_{-32}$ &
$^{+9}_{-9}$ & $^{+7}_{-7}$ & $^{+169}_{-141}$ & $^{+47\%}_{-39\%}$ \\  \hline
$164$ & 340 & $^{+72}_{-72}$ & $^{+28}_{-30}$ & $^{+44}_{-41}$ & $^{+36}_{-31}$ &
$^{+7}_{-7}$ & $^{+7}_{-7}$ & $^{+159}_{-133}$ & $^{+47\%}_{-39\%}$ \\  \hline
$165$ & 333 & $^{+70}_{-71}$ & $^{+28}_{-29}$ & $^{+44}_{-41}$ & $^{+35}_{-30}$ &
$^{+7}_{-7}$ & $^{+7}_{-7}$ & $^{+156}_{-130}$ & $^{+47\%}_{-39\%}$ \\  \hline 
$166$ & 324 & $^{+69}_{-69}$ & $^{+27}_{-29}$ & $^{+43}_{-40}$ & $^{+34}_{-29}$ &
$^{+6}_{-6}$ & $^{+7}_{-7}$ & $^{+151}_{-126}$ & $^{+47\%}_{-39\%}$ \\  \hline
$168$ & 310 & $^{+65}_{-66}$ & $^{+26}_{-28}$ & $^{+41}_{-38}$ & $^{+33}_{-28}$ &
$^{+5}_{-5}$ & $^{+7}_{-7}$ & $^{+143}_{-119}$ & $^{+46\%}_{-38\%}$ \\  \hline
$170$ & 297 & $^{+63}_{-63}$ & $^{+25}_{-27}$ & $^{+40}_{-37}$ & $^{+32}_{-27}$ &
$^{+4}_{-4}$ & $^{+6}_{-6}$ & $^{+137}_{-114}$ & $^{+46\%}_{-38\%}$ \\  \hline
$175$ & 267 & $^{+56}_{-57}$ & $^{+23}_{-25}$ & $^{+37}_{-33}$ & $^{+29}_{-25}$ &
$^{+3}_{-3}$ & $^{+5}_{-5}$ & $^{+123}_{-102}$ & $^{+46\%}_{-38\%}$ \\  \hline 
$180$ & 240 & $^{+50}_{-51}$ & $^{+22}_{-23}$ & $^{+34}_{-31}$ & $^{+26}_{-22}$ &
$^{+1}_{-1}$ & $^{+5}_{-5}$ & $^{+109}_{-90}$ & $^{+45\%}_{-38\%}$ \\  \hline
$185$ & 217 & $^{+45}_{-46}$ & $^{+20}_{-21}$ & $^{+31}_{-28}$ & $^{+24}_{-20}$ &
$^{+1}_{-1}$ & $^{+5}_{-5}$ & $^{+99}_{-82}$ & $^{+46\%}_{-38\%}$ \\  \hline 
$190$ & 196 & $^{+41}_{-42}$ & $^{+18}_{-19}$ & $^{+28}_{-25}$ & $^{+22}_{-18}$ &
$^{+2}_{-2}$ & $^{+4}_{-4}$ & $^{+91}_{-75}$ & $^{+46\%}_{-38\%}$ \\  \hline
$195$ & 178 & $^{+37}_{-38}$ & $^{+17}_{-18}$ & $^{+26}_{-23}$ & $^{+20}_{-17}$ &
$^{+2}_{-2}$ & $^{+3}_{-3}$ & $^{+83}_{-69}$ & $^{+47\%}_{-39\%}$ \\  \hline 
$200$ & 162 & $^{+33}_{-35}$ & $^{+16}_{-17}$ & $^{+25}_{-22}$ & $^{+19}_{-15}$ &
$^{+2}_{-2}$ & $^{+3}_{-3}$ & $^{+77}_{-63}$ & $^{+47\%}_{-39\%}$ \\  \hline
\end{tabular}  
\caption{The NNLO total Higgs production cross sections in the $\protect{gg\to 
H}$  process at the Tevatron (in fb) for given Higgs mass values (in GeV) with the 
corresponding uncertainties from the various sources discussed in section 3, 
as well as the total uncertainty when all errors are added using the procedure
described in the text. } 
\vspace*{-1mm}
}

   One then obtains the maximal and minimal values of the  cross
section when scale, PDF and $\alpha_s$ (both experimental and theoretical)
uncertainties  are included, 
\beq 
\sigma_{\rm max}^{\rm \mu+PDF+\alpha_s} &=& (\sigma_0 + \Delta
\sigma^+_\mu)+ \Delta  (\sigma_0 + \Delta \sigma^+_\mu)^+_{\rm PDF+
 \alpha_s^{\rm exp}+ \alpha_s^{\rm th}} 
\, , \nonumber \\
\sigma_{\rm min}^{\rm \mu+PDF+\alpha_s} &=& (\sigma_0 - \Delta \sigma^-_\mu)-
\Delta  (\sigma_0 - \Delta \sigma^-_\mu)^-_{\rm PDF+ \alpha_s^{\rm exp}
+ \alpha_s^{\rm th}}   \, . 
\label{combined}
\eeq 
To these new maximal and minimal cross sections, one should then add  the  much
smaller errors originating from the other sources such as,  in the case of the 
$gg\to H$  process, those due to the missing $b$--quark loop and the  mixed
QCD--electroweak corrections at NNLO.  This last addition can be done linearly 
as the errors from the use of the effective theory approach are purely
theoretical ones and do not  depend on the scale choice in practice\footnote{In
the case of the $b$--loop contribution, the $K$--factor when  varying the scale
from the central value $M_H$ to the values $\approx \frac13 M_H$ or $\approx
3M_H$.which  maximise and minimise the cross section, might be slightly
different and thus, the error will not be exactly that given in Table 2.
However, since the entire effect is very small, we will ignore this tiny
complication here.}.

The two last columns of Table 2 display the maximal and minimal deviations of
the $gg \to H$ cross section at the Tevatron when all errors are added, as 
well as the percentage deviations  of the cross section from the central
value. We should note that the actual PDF+$\alpha_s$ error and the error from
the use of the  effective theory approach   are different  from those of Table
2, which are given for the best value of the cross section, obtained for the
central scale choice $\mu_F=\mu_R=M_H$; nevertheless, the relative or 
percentage errors are approximately the same for $\sigma_0$ and $\sigma_0 \pm
\Delta \sigma^\pm_\mu$.

One observes from Table 2 that when all theoretical errors are combined, there
is a large variation of the $gg\to H$ cross section. The percentage total error
on the cross section is approximately the same in the entire Higgs mass range
that is indicated and is significant, the lower and upper values being $\approx
40\%$ smaller or $\approx 50\%$ larger than the central value. For $M_H=160$
GeV for instance,  one obtains a spread from the central value $\sigma_0= 374$
fb which amounts to  $\sigma_{\rm max}={552}$ fb and $\sigma_{\rm min}={225}$
fb, a spread that leads to a percentage error of $\Delta \sigma_0 / \sigma_0 
=-39.7\%$ and $+ {47.6}\%$.   

This is again summarized in Fig.~11, where the total uncertainty band obtained
in our analysis is confronted to the uncertainty band that one obtains when
adding in quadrature the scale uncertainty for $\frac12 M_H \le  \mu_R, \mu_F
\le 2 M_H$ and the PDF error only (without the errors on  $\alpha_s$) as
assumed in the CDF/D0 analysis. Furthermore, in the latter case, we use the
resumed NNLL cross sections given in Ref.~\cite{ggH-FG} which is $\sim 15\%$
higher than the cross section that we obtain when including the higher order
contributions only to NNLO  and has a milder scale variation. As can be seen,
the difference between the two uncertainty bands is striking. In fact, even the
lower value of the cross section in the NNLL approach, including the scale and
PDF errors when combined in quadrature, only touches the central value of our
NNLO result. For $M_H=160$ GeV, the lower value of the cross section, when all
errors are included, is $\approx 40\%$ smaller than the central value at NNLO
and $\approx 50\%$ smaller than the NNLL cross section adopted in
Ref.~\cite{Tevatron} as a normalisation.

\begin{figure}[!h]
\vspace*{1mm}
\begin{center}
\epsfig{file=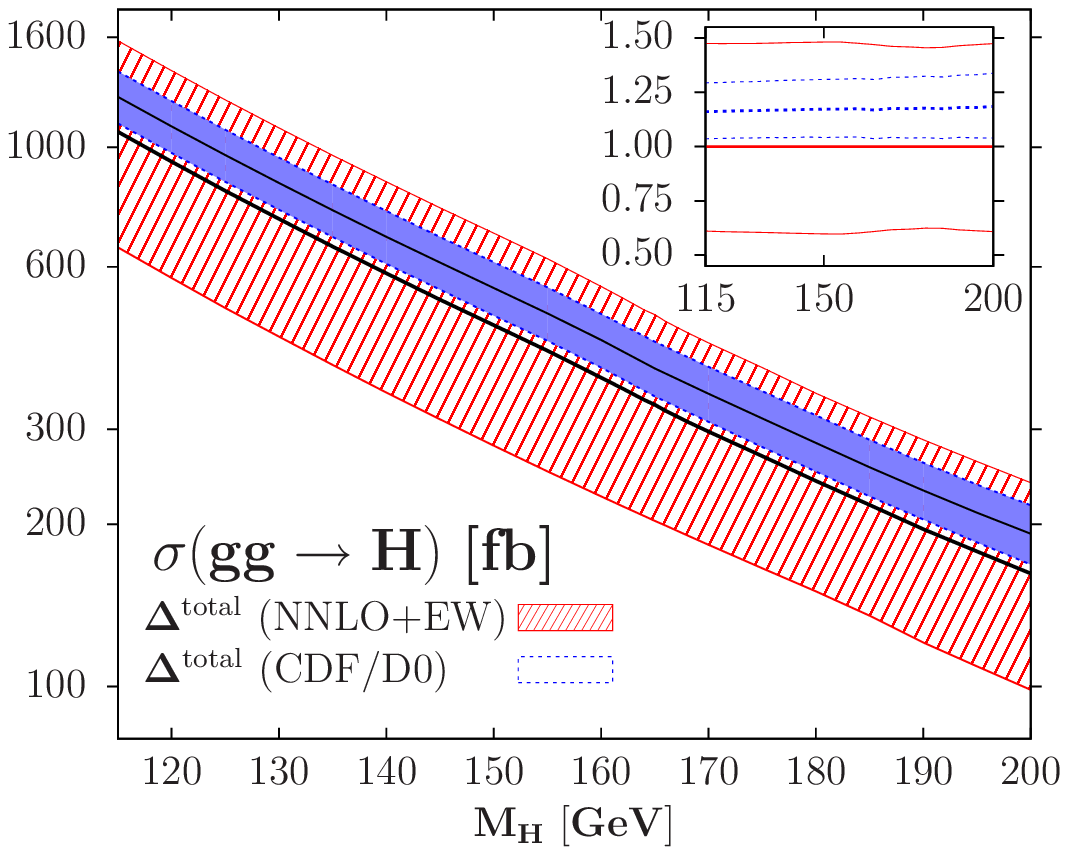,width=11cm}
\end{center}
\vspace*{-5mm}
\caption[]{The production cross section $\sigma(gg\to H)$ at NNLO at
the Tevatron with the uncertainty band when all the errors are added
using our procedure (last columns of Table 2). It is compared to $\sigma(gg 
\to H)$ at NNLL when the scale and PDF errors given in Ref.~\cite{ggH-FG} are 
added in quadrature. In the insert the relative deviations are shown when 
the central values are normalized to $\sigma^{\rm NNLO+EW}$.} 
\end{figure}

\TABLE[!h]{\small%
\let\lbr\{\def\{{\char'173}%
\let\rbr\}\def\}{\char'175}%
\renewcommand{\arraystretch}{1.66}
\begin{tabular}{|c|cc||c|ccc|cc|}\hline 
$~M_H~$ & $~\sigma_{HW}$  & $\sigma_{HZ}$~ &~~~scale~~~&PDF&PDF+$\alpha_s
^{\rm exp}$ & $~~\alpha_s^{\rm th}~~$ &  total & \% total \\ \hline 
$115$ & 174.5 & 103.9 & $^{+1.3}_{-1.6}$ & $^{+10.5}_{-9.1}$ & $^{+10.7}_{-10.7}$ &
$^{+1.3}_{-0.9}$ & $^{+12.1}_{-12.3}$ & $^{+7\%}_{-7\%}$ \\  \hline 
$120$ & 150.1 & 90.2 & $^{+1.1}_{-1.4}$ & $^{+9.2}_{-8.1}$ & $^{+9.6}_{-9.4}$ &
$^{+1.2}_{-0.9}$ & $^{+10.7}_{-10.9}$ & $^{+7\%}_{-7\%}$ \\  \hline
$125$ & 129.5 & 78.5 & $^{+0.9}_{-1.3}$ & $^{+7.5}_{-6.8}$ & $^{+8.6}_{-8.7}$ &
$^{+1.1}_{-0.8}$ & $^{+9.6}_{-10.0}$ & $^{+7\%}_{-8\%}$ \\  \hline  
$130$ & 112.0 & 68.5 & $^{+0.8}_{-1.1}$ & $^{+6.8}_{-6.4}$ & $^{+7.2}_{-7.5}$ &
$^{+1.1}_{-0.8}$ & $^{+8.0}_{-8.6}$ & $^{+7\%}_{-8\%}$ \\  \hline 
$135$ & 97.2 & 60.0 & $^{+0.7}_{-1.0}$ & $^{+5.6}_{-5.5}$ & $^{+6.7}_{-6.6}$ &
$^{+1.0}_{-0.7}$ & $^{+7.4}_{-7.6}$ & $^{+8\%}_{-8\%}$ \\  \hline 
$140$ & 84.6 & 52.7 & $^{+0.6}_{-0.9}$ & $^{+5.6}_{-4.5}$ & $^{+5.8}_{-5.7}$ &
$^{+0.9}_{-0.7}$ & $^{+6.5}_{-6.6}$ & $^{+8\%}_{-8\%}$ \\  \hline 
$145$ & 73.7 & 46.3 & $^{+0.5}_{-0.8}$ & $^{+4.4}_{-4.1}$ & $^{+5.4}_{-5.2}$ &
$^{+0.9}_{-0.7}$ & $^{+5.9}_{-6.0}$ & $^{+8\%}_{-8\%}$ \\  \hline 
$150$ & 64.4 & 40.8 & $^{+0.5}_{-0.7}$ & $^{+4.2}_{-3.9}$ & $^{+4.4}_{-4.3}$ &
$^{+0.8}_{-0.6}$ & $^{+5.0}_{-5.0}$ & $^{+8\%}_{-8\%}$ \\  \hline 
$155$ & 56.2 & 35.9 & $^{+0.4}_{-0.6}$ & $^{+3.4}_{-3.1}$ & $^{+4.2}_{-4.1}$ &
$^{+0.7}_{-0.6}$ & $^{+4.6}_{-4.7}$ & $^{+8\%}_{-8\%}$ \\  \hline 
$160$ & 48.5 & 31.4 & $^{+0.4}_{-0.6}$ & $^{+3.3}_{-3.0}$ & $^{+3.6}_{-3.3}$ &
$^{+0.7}_{-0.5}$ & $^{+4.1}_{-4.0}$ & $^{+8\%}_{-8\%}$ \\  \hline 
$162$ & 47.0 & 30.6 & $^{+0.4}_{-0.5}$ & $^{+3.4}_{-2.8}$ & $^{+3.5}_{-3.3}$ &
$^{+0.7}_{-0.5}$ & $^{+3.9}_{-3.8}$ & $^{+8\%}_{-8\%}$ \\  \hline 
$164$ & 44.7 & 29.1 & $^{+0.3}_{-0.5}$ & $^{+3.1}_{-2.7}$ & $^{+3.4}_{-3.4}$ &
$^{+0.6}_{-0.5}$ & $^{+3.7}_{-3.9}$ & $^{+8\%}_{-9\%}$ \\  \hline 
$165$ & 43.6 & 28.4 & $^{+0.3}_{-0.5}$ & $^{+2.8}_{-2.4}$ & $^{+3.4}_{-3.3}$ &
$^{+0.6}_{-0.5}$ & $^{+3.8}_{-3.8}$ & $^{+8\%}_{-8\%}$ \\  \hline 
$166$ & 42.5 & 27.8 & $^{+0.3}_{-0.5}$ & $^{+3.0}_{-2.6}$ & $^{+3.1}_{-3.0}$ &
$^{+0.6}_{-0.5}$ & $^{+3.4}_{-3.5}$ & $^{+8\%}_{-8\%}$ \\  \hline 
$168$ & 40.4 & 26.5 & $^{+0.3}_{-0.5}$ & $^{+2.8}_{-2.4}$ & $^{+3.1}_{-2.9}$ &
$^{+0.6}_{-0.5}$ & $^{+3.4}_{-3.4}$ & $^{+9\%}_{-8\%}$ \\  \hline 
$170$ & 38.5 & 25.3 & $^{+0.3}_{-0.4}$ & $^{+2.9}_{-2.2}$ & $^{+3.0}_{-2.7}$ &
$^{+0.6}_{-0.5}$ & $^{+3.3}_{-3.1}$ & $^{+9\%}_{-8\%}$ \\  \hline 
$175$ & 34.0 & 22.5 & $^{+0.3}_{-0.4}$ & $^{+2.2}_{-1.9}$ & $^{+2.7}_{-2.6}$ &
$^{+0.5}_{-0.4}$ & $^{+3.0}_{-3.0}$ & $^{+9\%}_{-9\%}$ \\  \hline 
$180$ & 30.1 & 20.0 & $^{+0.2}_{-0.4}$ & $^{+2.1}_{-1.8}$ & $^{+2.2}_{-2.2}$ &
$^{+0.5}_{-0.4}$ & $^{+2.5}_{-2.6}$ & $^{+8\%}_{-9\%}$ \\  \hline 
$185$ & 26.9 & 17.9 & $^{+0.2}_{-0.3}$ & $^{+1.8}_{-1.5}$ & $^{+2.1}_{-2.1}$ &
$^{+0.5}_{-0.4}$ & $^{+2.3}_{-2.4}$ & $^{+9\%}_{-9\%}$ \\  \hline 
$190$ & 24.0 & 16.1 & $^{+0.2}_{-0.3}$ & $^{+1.6}_{-1.6}$ & $^{+1.8}_{-1.8}$ &
$^{+0.4}_{-0.3}$ & $^{+2.1}_{-2.1}$ & $^{+9\%}_{-9\%}$ \\  \hline 
$195$ & 21.4 & 14.4 & $^{+0.2}_{-0.3}$ & $^{+1.3}_{-1.2}$ & $^{+1.8}_{-1.7}$ &
$^{+0.4}_{-0.3}$ & $^{+2.1}_{-2.0}$ & $^{+10\%}_{-10\%}$ \\  \hline 
$200$ & 19.1 & 13.0 & $^{+0.2}_{-0.2}$ & $^{+1.4}_{-1.2}$ & $^{+1.5}_{-1.4}$ &
$^{+0.4}_{-0.3}$ & $^{+1.8}_{-1.7}$ & $^{+9\%}_{-9\%}$ \\  \hline 
\end{tabular} 
\vspace*{2mm}
\caption{The central values of the cross sections for the $p \bar p\to WH$ 
and $ZH$ processes at the Tevatron  (in fb) for given Higgs mass values (in 
GeV) with, in the case of the $WH$ channel,  the uncertainties from  scale 
variation, PDF, PDF+$\Delta^{\rm exp} \alpha_s$ and $\Delta^{\rm th}\alpha_s$, as 
well as the total uncertainty when all errors are added using the procedure 
described in the text.}
\vspace*{-1mm}
}

We thus believe that the CDF/D0 combined analysis which rules out the 162--166
GeV mass range for the SM Higgs boson on the basis of the $gg \to H \to \ell
\ell \nu \nu+X$ process, which is the most (if not the only) relevant one in
this specific mass range at the Tevatron, has largely underestimated the
theoretical errors on the Higgs production cross section. In fact, even if
the scale uncertainty were taken to be that resulting from a variation in the
usual domain $\frac12 M_H \le \mu_F, \mu_R \le 2M_H$ or the errors from the use
of the effective approach at NNLO were ignored, the total uncertainty would 
have been of the order of $\approx 35\%$, i.e three times larger than the error
assumed in the CDF/D0 analysis.

Turning to the Higgs--strahlung processes, and similarly to the $gg \to H$
case, we display in Table 3 the central values of the cross sections for $p
\bar p \to WH$ and $p \bar p \to ZH$ at the Tevatron, evaluated at scales
$\mu_R=\mu_F=M_{HV}$ with the MSTW set of PDFs (second and third columns). In
the remaining columns, we specialize in the $WH$ channel and display the
errors from the scale variation (with $\kappa=2$), the PDF, mixed
PDF+$\Delta^{\rm exp} \alpha_s$ and PDF+$\Delta^{\rm exp} \alpha_s$+$\Delta^
{\rm th} \alpha_s$ uncertainties in the MSTW scheme. In the last columns, we
give the total error and its percentage; this percentage error is, to a very
good approximation, the same  in the $p \bar p \to ZH$ channel. In contrast to
the $gg\to H$ mechanism, since the errors due to scale variation are rather
moderate in this case, there is no large difference between the central cross
section $\sigma_0$ and the  cross sections $\sigma_0 \pm \Delta \sigma^\pm_\mu$
and, hence, the PDF,  PDF+$\Delta^{\rm exp} \alpha_s$ and PDF+$\Delta^{\rm
exp} \alpha_s$+$\Delta^ {\rm th} \alpha_s$  errors  on  $\sigma^0$ are, to a
good  approximation, the same as the errors on $\sigma_0 \pm \Delta
\sigma^\pm_\mu$  displayed in Table 3.  

 The total uncertainty is once more summarized in Fig.~12, where the cross
sections for $WH$ and  $ZH$ associated production at the Tevatron, together
with the total uncertainty bands (in absolute values in the main frame and in
percentage in the insert), are displayed as a function of the Higgs mass.    As
can be seen, the total error  on the cross sections in the Higgs--strahlung 
processes is about $\pm 9\%$ in the entire Higgs mass range, possibly 1\% to
2\% smaller for low $M_H$ values and $\sim 1\%$ larger for high $M_H$ values. 
Thus, the theoretical errors are much smaller than in the case of the $gg \to
H$ process and the cross sections for the Higgs--strahlung processes are well
under control. Nevertheless, the total uncertainty obtained in our analysis  is
almost twice as large  as the total 5\% uncertainty assumed by the CDF and D0
collaborations in their combined analysis of this channel \cite{Tevatron}.

\begin{figure}[h]
\vspace*{-1mm}
\begin{center}
\epsfig{file=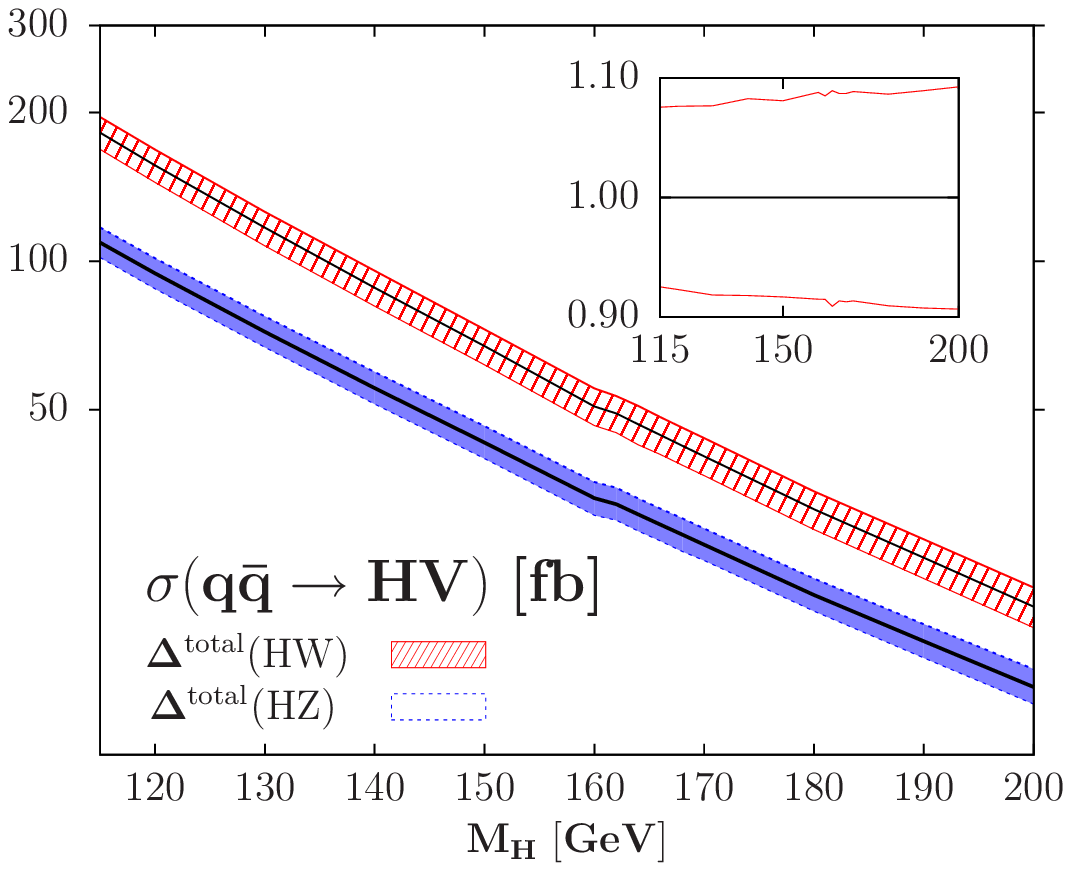,width=11.cm}
\end{center}
\vspace*{-6mm}
\caption[]{The production cross section $\sigma(p \bar p \to WH)$ and 
$\sigma(p \bar p \to ZH)$ at  NNLO in QCD and electroweak NLO at the Tevatron
evaluated with the MSTW set  of PDFs, together with the uncertainty bands when
all the theoretical errors  are added. In the insert, the relative deviations
from the central MSTW value are shown in the case of $\sigma(p\bar p \to WH)$.} 
\vspace*{-1mm}
\end{figure}

Before closing this section, let us mention that the uncertainties in the 
Higgs--strahlung processes can be significantly reduced by using the  Drell--Yan
processes of massive gauge boson production as standard candles; a suggestion
first made in  Ref.~\cite{DY-normal}. Indeed, normalizing the cross sections of
associated $WH$ and $ZH$ production to the cross sections of single $W$ and $Z$
production, respectively, allows for a cancellation of several experimental
errors such as the error on the luminosity measurement, as well as the partial 
cancellation (since the scales that are involved in the $p\bar p \to V$ and
$HV$  processes are different) theoretical errors such as those due to the
PDFs, $\alpha_s$  and the higher order radiative corrections.

\section{Conclusion}

In the first part of this paper, we have evaluated the production cross
sections of the Standard Model Higgs boson at the Tevatron, focusing on the two
main   channels: the gluon--gluon fusion $gg \to H$ mechanism that dominates in
the  high Higgs mass range and the Higgs--strahlung processes $q\bar q \to VH$
with $V=W,Z$, which are the most important ones in the lower Higgs mass range. 
In the determination of the cross sections, we have included all the available
and relevant higher order corrections in perturbation theory, in particular,
the QCD corrections up to NNLO and the one--loop electroweak radiative
corrections. We have then provided up--to--date central values of the cross
sections for the the entire Higgs mass range that is relevant at the Tevatron.
While this update has been performed for the $gg\to H$ mechanism in several
recent analyses, it was missing in the case of the Higgs--strahlung processes.

The second part of the paper addresses the important issue of the theoretical
uncertainties that affect the predicted cross sections. We have first discussed the
scale uncertainties which are usually viewed as a measure of the unknown higher
order contributions. Because the calculated QCD corrections are extremely large
in the $gg\to H$ process,  we point out that the domain of variation of the
renormalisation and factorisation scales that is usually adopted in the
literature should be extended. We adopt a criterion that allows for a more
reasonable or conservative estimate of this variation domain:  the range of
variation of the scales at NNLO, should be the one which allows to the scale
uncertainty band of the NLO cross section to include the NNLO contributions.
Applying this  criterion to the NNLO $gg\to H$ cross section and adopting a
central scale $\mu_0=M_H$,   we obtain a scale uncertainty  of the order of
$\pm 20\%$, i.e. slightly larger than the $\approx \pm 15\%$ uncertainty that
is usually assumed. This larger error would at least account for the 20--30\%
discrepancy between the QCD corrections to the inclusive cross section that is
used as a normalisation and the cross section with the basic kinematical cuts
applied  in the experimental analyses.

A second source of uncertainties in the $gg\to H$ cross section originates  
from the use of the effective theory approach that allows to considerably
simplify the calculation of the NNLO contributions, an approach in which the
masses of the loop particles that generate the $Hgg$ vertex  are assumed to be
much larger than the Higgs mass. We show that the missing NNLO contribution of
the $b$--quark loop where the limit $M_H \ll m_b$ cannot be applied (together
with the definition of the $b$--quark mass), and the approximation $M_H \ll
M_W$ used in the three--loop mixed QCD--electroweak NNLO radiative corrections,
might lead to a few percent error on the total $gg \to H$ cross section in each
case.

A third source of theoretical errors  is due to the parton distribution
functions and the errors associated  to  the strong coupling constant. 
Considering not only the MSTW scheme as usually done, but also the  CTEQ and
ABKM schemes, we recall that while the PDF errors are relatively small within a
given scheme, the central values can be widely different. This is particularly 
true in the case of the $gg \to H$ cross section, where the central values in
the MSTW/CTEQ and ABKM schemes differ by about 25\%. Only  when the
experimental as well as the theoretical errors on $\alpha_s$ are accounted for
that one obtains results that are consistent when using the MSTW/CTEQ and ABKM 
schemes. In the MSTW scheme, using a recently released set--up which provides a
simultaneous access to the PDF and $\Delta^{\rm exp}\alpha_s$ errors as well as
a way to estimate the $\Delta^{\rm th}\alpha_s$ error, one finds a $\approx
15\%$ uncertainty on $\sigma(gg \to H)$, that is, at least  a factor of two
larger than the uncertainty due to  the PDFs alone that is usually considered
as  the total PDF error

We have then proposed a simple procedure to combine these various theoretical 
errors. The main idea  of this procedure is  to evaluate directly the
PDF+$\Delta^{\rm exp}\alpha_s$+$\Delta^{\rm th}\alpha_s$ error, as well as the
significantly smaller errors due to the use of the effective approach in the
$gg \to H$ process at NNLO, on the maximal and minimal values of the cross
sections that one obtains when varying the renormalisation and factorisation
scales in the chosen domain.  

Adopting this  approach, one arrives at a total uncertainty of $\approx -40\%$
and $\approx +50\%$ for the central value of the $gg \to H$ cross section at
the Tevatron, a much larger error than the $\approx 10\%$ uncertainty that is
usually assumed\footnote{We note that it would be interesting to study the
impact of these theoretical uncertainties on the $gg \to H$ cross sections for
Higgs production at the LHC, not only for the discovery of the particle, but
also for the measurement of its couplings to fermions and gauge bosons
\cite{Higgs-cpg}, which is another crucial issue in this context. A preliminary
analysis shows that at $\sqrt s= 14$ TeV, the total error that one obtains on
the  NNLO total production cross section is of the order of 25\% for $M_H
\approx 160$ GeV, i.e. much less than at the Tevatron. The main reason is that
the PDF+$\alpha_s$ uncertainties are slightly smaller than those obtained for
the Tevatron, while the scale uncertainty (in which one needs only the more
reasonable  factor $\kappa=2$) is reduced, $\lsim 15\%$, a mere consequence of
the fact that the $K$--factors are more moderate, $K_{\rm NNLO} \approx 2$ at
the LHC instead of $K_{\rm NNLO}  \approx 3$ at the Tevatron.   More details
will be given elsewhere \cite{Julien}.}.  Hence, the number of signal events
from  the $gg \to H$ process with the subsequent Higgs decay $H \to WW \to \ell
\ell \nu \nu$, i.e. the main  (if not the only  relevant) Higgs channel at the
Tevatron in the Higgs mass range 150 GeV $\lsim M_H \lsim 180$ GeV, might be a
factor of two smaller than what has been assumed by  the CDF and D0 
collaborations in their recent analysis which excluded the Higgs mass range
between 162 and 166 GeV. We thus believe  that this analysis should be
reconsidered in the light of these larger theoretical uncertainties in the
signal cross sections\footnote{This is without considering  the uncertainties
in the background cross sections.  It would be indeed interesting to apply our
procedure for the evaluation of  the scale variation, the PDF+$\alpha_s$ errors
and for their combination, to the major expected backgrounds of the Higgs
signal, namely Drell--Yan, top quark pair and gauge boson pair production. This
is, however, beyond the scope of the present paper.}. 

Of course, one can view the results presented in this paper  with a more
optimistic perspective: since the uncertainties in the $gg\to H$ process are so
large, the cross section might well be closer to its upper limit which is
$\approx 50\%$ higher than the central value. In this lucky situation, the
sensitivities of the CDF and D0 collaborations would be significantly increased
and if the Higgs boson happens to have a mass in the range $M_H \approx
160$--170 GeV,  some evidence for the particle at the Tevatron might soon show
up.

Finally, in the case of the Higgs--strahlung processes, the cross sections are 
much more under control, the main reason being due to the fact that the QCD
corrections are moderate. The scale uncertainties are at percent level for the
narrow domain chosen for the scale variation (within a factor of two from the
central scale), while the PDF uncertainties and the associated   uncertainties
due to the experimental and theoretical errors on $\alpha_s$  are much smaller
than in the $gg \to H$ case. The total estimated theoretical  error on the
Higgs--strahlung cross sections, $\approx 10\%$, is nevertheless almost twice
as large as the error assumed by the CDF and D0 collaborations.  

\subsection*{Acknowledgments} 

This work is supported by the European network HEPTOOLS. JB thanks the CERN TH
group for the hospitality extended to him during this work.  Discussions with 
Babis Anastasiou, Michael Dittmar, Davide Gerbaudo, Massimiliano Grazzini, Wade
Fisher, Robert Harlander, Vajravelu Ravindran, Peter Skands, Michael Spira and
Peter Uwer are gratefully acknowledged.   Special thanks go to Michael Spira
for his critical comments on an earlier version of the manuscript.

%\newpage

\end{document}